\documentclass[aps,pra,twocolumn,superscriptaddress,showpacs,showkeys,amsmath,amssymb]{revtex4}

\usepackage{amsfonts}
\usepackage{amssymb,amsmath}
\usepackage{mathrsfs}
\usepackage{latexsym}
\usepackage{amsmath}
\usepackage[cp1251]{inputenc}
\usepackage{graphicx}
\usepackage{dcolumn}
\usepackage{bm}
\usepackage{color}

\RequirePackage{ifthen}
\RequirePackage[pdfstartview=FitH]{hyperref}
\begin{document}
	
\title{Quantum dynamics of spinful impurity in ideal Bose gas}
\author{O.~Hryhorchak\footnote{e-mail: hrorest@gmail.com}}
\affiliation{Professor Ivan Vakarchuk Department for Theoretical Physics, Ivan Franko National University of Lviv, 12 Drahomanov Street, Lviv, Ukraine}
\author{V.~Pastukhov\footnote{e-mail: volodyapastukhov@gmail.com}}
\affiliation{Professor Ivan Vakarchuk Department for Theoretical Physics, Ivan Franko National University of Lviv, 12 Drahomanov Street, Lviv, Ukraine}

\date{\today}

\pacs{67.85.-d}
	
\keywords{spin in bosonic bath, spin-dependent interaction, Rabi oscillations}
	
\begin{abstract}
We discuss the quench dynamics of an isolated system composed of a single spinful impurity in the transverse Rabi field and bath of non-interacting three- and two-dimensional bosons. Specifically, we consider the evolution of bosons and a spin-$\frac{1}{2}$ particle, initially prepared in a Bose-Einstein condensate state and a magnetic ground state, respectively, with the spin-dependent contact boson-impurity interaction switched on. Applying an original mean-field-like approximation, which naturally reflects the statistical effects of the bosonic bath, we calculate time-dependent components of the average impurity spin and the overlap of the wave function between initial and arbitrary time moments. A key prediction is a substantial speed-up in the decoherence (thermalization) dynamics of the spin degree of freedom compared to results obtained with the Chevy-like ansatz.
\end{abstract}
	
\maketitle
\section{Introduction}
A single impurity in a bosonic medium is a paradigmatic model in many physical contexts, including solid-state physics, quantum optics, ultracold atomic gases, etc. In the latter example, one typically deals with a bath formed by an ensemble of atoms, rather than with quasiparticles or electromagnetic field quanta (phonons and photons, respectively). This substantially enriches the physics of such systems, known as Bose polarons, with nontrivial few- and many-body effects. At the same time, ultracold-gas settings provide precise control over system parameters, such as the system's composition, the magnitude of boson-impurity interactions, and the internal states of the impurity. In the last decade, the properties of impurities in Bose-Einstein condensates have been extensively discussed both theoretically and experimentally (see recent reviews \cite{Mistakidis_2023, Grusdt_2025, Massignan_2026}). 

A typical Bose-polaron setup assumes a dilute Bose gas interacting with the impurity. Macroscopic occupation of the lowest energy level by bosons identifies \cite{Grusdt_2016, Grusdt_2015, Shchadilova_2016} the system as being very similar to Fr\"ohlich polaron \cite{Frohlich_1954}. However, it is only true for weak boson-impurity couplings, and an increase in interaction makes this picture oversimplified \cite{Grusdt_2017, Kain_2018, Ichmoukhamedov_2019, Christ_2025}, requiring inclusion of the two-phonon terms to the original Fr\"ohlich Hamiltonian. Such a model complication necessarily leads to the formation of boson-impurity bound states \cite{Rath_2013, Li_2014, Pena_Ardila_2015, Alhyder_2026} that can be observed experimentally \cite{Hu_2016, Jorgensen_2016, Yan_2020, Skou_2022, Etrych_2025}. Bose statistics of host atoms support the formation of higher few-body bound states \cite{Yoshida_2018, Shi_2018, Levinsen_2021}. The simplest ones are composed of an impurity and two bosonic atoms and are known \cite{Efimov_1970} to yield an infinite number of three-body bound states in vacuum as the boson-impurity $s$-wave scattering length diverges. The impact of the Efimov effect on the impurity spectrum in the Bose-polaron setting was studied in Refs.~\cite{Levinsen_2015, Sun_2017,  Yoshida_2018_2, Blume_2019}. On the other hand, the bosonic cloud accompanying the impurity that the Gaussian-state ansatz can account for \cite{Christianen_2022, Christianen_2022_2} shifts the position of the Efimov resonance. The inclusion of the wave-function non-Gaussianity predicts \cite{Mostaan_2025} a discrete set of metastable many-body bound states between attractive and repulsive polaron branches. However, a rich physics of many-body resonances in strongly coupled Bose polarons requires new theoretical approaches, with experiments underway \cite{Robens_2026}.

The described variety of scenarios for Bose-polaron behavior mostly concerns equilibrium states. In typical ultracold-atom experiments, the system is first prepared in a non-interacting or weakly interacting regime, after which the impurity-boson scattering length, and consequently the interaction strength, is tuned via a magnetic Feshbach resonance. Therefore, the true ground state of the system is rarely achievable, and the quench dynamics take precedence. The dynamics of the Bose-polaron formation was observed experimentally \cite{Skou_2021} and studied theoretically \cite{Shchadilova_2016_2, Nielsen_2019, Dzsotjan_2020} beyond the Fr\"ohlich paradigm. For time-dependent calculations, standard perturbative approaches are typically inapplicable, and one needs to exploit the summation of infinite diagrammatic series \cite{Volosniev_2015} or develop sophisticated variational methods \cite{Field_2020, Pena_Ardila_2021}. Exact results are only possible \cite{Drescher_2021} in the limit of a static impurity immersed in an ideal Bose gas. Assuming a nonzero impurity spin and that the impurity interacts with bosons differently depending on its internal state enables radio-frequency spectroscopy of the system. The latter provides an individual quantum degree of freedom for probing a many-body environment. The appropriate spectra for a mobile impurity in a dilute Bose gas were obtained in Ref.~\cite{Shashi_2014}. There is another experimentally accessible way \cite{Adam_2022} to generate polaron dynamics by fine-tuning the boson-impurity scattering length for one internal state of impurity to resonant magnitude. At the same time, the coupling to another spin state is left almost unchanged, causing nontrivial wave-function evolution \cite{Mistakidis_2019, Hryhorchak_2026}. This is also a simplified model for Rydberg impurity \cite{Schmidt_2016, Camargo_2018, Durst_2024} in Bose-Einstein condensates. The situation becomes more intriguing when the transverse field acts on the spin, introducing additional quantum dynamics known as Rabi oscillations even in the absence of a bosonic bath. Recently, this setup was discussed \cite{Mulkerin_2024, Wasak_2024, Bleu_2025, Vivanco_2025} in the context of the Fermi-polaron problem and for a dilute Bose gas \cite{Liu_2026}, including in the wave function up to a single excitation above the Bogoliubov ground state. Here, we adopt an original mean-field-like approach in which all bosons are allowed to evolve simultaneously through time-dependent single-particle orbitals. This retains the collective response of the Bose condensate while avoiding an explicit truncation of the number of excited bosons. And this is done for three- and two-dimensional geometries, revealing an impact of spatial dimension on the observables.

\section{Model and problem statement}
We consider a system of an ideal Bose gas interacting with a spinful impurity. The system is loaded in the $d$-dimensional large volume $L^d$ with periodic boundary conditions. The boson-impurity interaction is assumed to be contact and spin-dependent, with the only nonzero coupling constants $g_{\uparrow}$ and $g_{\downarrow}$, respectively. We also assume the transverse magnetic field (Rabi coupling) acting on the spin. This combination of spin-up and spin-down interactions, together with the Rabi field, makes the quantum spin dynamics nontrivial. Before we study the collective behavior of a macroscopic number $N$ of bosons, which can be done only approximately, it is instructive to consider the limit of a single Bose particle. 
\subsection{Single-boson limit}
The appropriate Hamiltonian with separated relative motion ${\bf r}$ and center-of-mass coordinates at zero total momentum set, reads:
\begin{align}\label{h}
	&h=h_0+\Phi, \\
    &h_0=\varepsilon-\Delta \sigma_x, \ \ \Phi=\delta_{\Lambda}({\bf r})\sum_{\sigma={\uparrow,\downarrow}}g_{\sigma,\Lambda}|\sigma\rangle\langle \sigma|,
\end{align}
where $\varepsilon=\frac{{\bf p}^2}{2m}$ and $\sigma_x=|\uparrow\rangle\langle \downarrow|+|\downarrow\rangle\langle \uparrow|$. Without loss of generality, the Rabi coupling $\Delta$ is taken to be real-valued. The projection operator $|\sigma\rangle\langle \sigma'|$ acts on the impurity spin variable. The boson-impurity reduced mass is denoted by $m$, and the $d$-dimensional $\delta$-function is regularized $\frac{1}{L^d}\sum_{|{\bf p}|\le \Lambda}e^{i{\bf p}{\bf r}}$ by the ultraviolet (UV) cutoff $\Lambda$. The bare couplings $g_{\sigma,\Lambda}$
\begin{subequations}
\begin{align}
    g^{-1}_{\sigma,\Lambda}=g^{-1}_{\sigma}-\frac{1}{L^d}\sum_{|{\bf p}|<\Lambda}\frac{1}{\varepsilon_{\bf p}}, \ \ (d>2), \label{eq:2.3a}\\
    g^{-1}_{\sigma,\Lambda}=-\frac{1}{L^2}\sum_{|{\bf p}|<\Lambda}\frac{1}{\varepsilon_{\bf p}+|\epsilon_{\sigma}|}, \ \ (d=2), \label{eq:2.3b}
\end{align}
\end{subequations}
should be adjusted appropriately \cite{Hryhorchak_2023} to compensate UV divergences. In $d> 2$ for positive `observable' couplings $g_{\sigma}$
\begin{eqnarray}\label{g_def}
g^{-1}_{\sigma}=-\frac{\Gamma(1-d/2)}{(2\pi)^{d/2}}m^{d/2}|\epsilon_{\sigma}|^{d/2-1},
\end{eqnarray}
a single $s$-wave bound state with energy $\epsilon_{\sigma}$ is guaranteed for each spin state of the impurity. The two-dimensional geometry always supports a bound state.

For our model, at least two examples of quench dynamics can be considered. In the first one, it is assumed that the system is initially prepared in a non-interacting state (all $g_{\sigma}=0$), and then the boson-impurity interaction is suddenly switched on. The second protocol suggests the system is in its lowest two-body bound state without the Rabi field, and then the Rabi field is turned on at $t=0$.  
Below, the macroscopic system is considered in the thermodynamic limit; therefore, only the first scenario is relevant.

A key quantity that will be important in further calculations for the first quench scenario is the matrix element of the evolution operator $\langle {\bf p}|e^{-ith}|{\bf p}'\rangle$ between relative-momentum states. This is still a $2\times 2$ unitary matrix in spin space. The time dependence of any system's state can be found as a linear combination involving these matrix elements. For later convenience, it is useful to introduce the matrix function with element $C_{{\bf p},{\bf p}'}(t)=-i\theta(t)\langle {\bf p}|e^{-ith}|{\bf p}'\rangle$ (here $\theta(t)$ is the Heaviside step function). By the Fourier transform
\begin{eqnarray}\label{G_omega}
\langle {\bf p}|\mathcal{G}(\omega_+)|{\bf p}'\rangle=\int_{-\infty}^{\infty}dt\,e^{it\omega_{+}}C_{{\bf p},{\bf p}'}(t),
\end{eqnarray}
(where $\omega_+=\omega+i0_+$ is the infinitesimally shifted in the complex plane frequency) one relates function $C_{{\bf p},{\bf p}'}(t)$ to the Green's function defined as follows
\begin{eqnarray}
\mathcal{G}(\omega)=\frac{1}{\omega-h}.
\end{eqnarray}
The latter can be conventionally rewritten in terms of that $\mathcal{G}^{(0)}(\omega)=\frac{1}{\omega-h_0}$ without interaction between translational and spin degrees of freedom, and the $2\times 2$ (because of spin degrees of freedom) $T$-matrix, $\mathcal{T}(\omega)$
\begin{eqnarray}
&&\mathcal{G}(\omega)=\mathcal{G}^{(0)}(\omega)+\mathcal{G}^{(0)}(\omega)\mathcal{T}(\omega)\mathcal{G}^{(0)}(\omega),\\
&&\mathcal{T}(\omega)=\Phi+\Phi\mathcal{G}^{(0)}(\omega)\mathcal{T}(\omega).
\end{eqnarray}
The contact character of the potential allows for explicit calculations \cite{Panochko_2021, Panochko_2022, Hryhorchak_2023_2} of the $T$-matrix element in momentum space, even in the case of spin-dependent interaction. The presence of a transverse field complicates the computations of $\mathcal{T}(\omega)$; however, the generic procedure can be adopted from Ref.~\cite{Hryhorchak_2026}. Particularly, for the simplified case $g_{\uparrow}\neq 0$ and $g_{\downarrow}= 0$, mainly discussed recently in context of fermionic \cite{Mulkerin_2024,Wasak_2024,Vivanco_2025} and bosonic \cite{Liu_2026} baths, the result for the $T$-matrix element is as follows:
\begin{eqnarray}
&&\langle {\bf p}|\mathcal{T}(\omega)|{\bf p}'\rangle=\frac{\tau(\omega)}{L^d}|\uparrow\rangle\langle \uparrow|,\label{Tau}\\
&&\tau^{-1}(\omega)=g^{-1}_{\uparrow,\Lambda}+\frac{1}{2}\Pi(\omega-\Delta)+\frac{1}{2}\Pi(\omega+\Delta),\nonumber
\end{eqnarray}
with $\tau(\omega)$ being a function (not a matrix) and short-hand notation $\Pi(\omega)=\frac{1}{L^d}\sum_{|{\bf p}|<\Lambda}\frac{1}{\varepsilon_{\bf p}-\omega}$ adopted. It is straightforward to verify that the function $\tau(\omega)$ remains finite in the $\Lambda\to \infty$ limit below $d=4$. With Eq.~\ref{Tau} calculated in arbitrary dimension, one can further obtain the evolution of the plane-wave state $e^{-ith}|{\bf p}\rangle$ by the inverse Fourier transform. For a single boson in a large volume $L^d$, the nontrivial term in the Green's function matrix element \ref{G_omega} is of order $1/L^d$. However, when multiplied by the number of Bose particles $N$, it determines the leading-order low-density contribution to observables.

\subsection{Many-body limit}
The macroscopic number of bosons requires the second-quantization methods to be utilized. For a finite impurity mass, the problem is complicated by the impurity's translational degrees of freedom. The standard tool in such cases is the use of the celebrated Lee-Low-Pines transformation \cite{LLP_1953}, which reduces a problem to a system of bath particles in the external field of a motionless impurity. However, if the impurity mass is finite, there is an additional term describing the kinematic interaction between host bosons. Keeping in mind variational calculations with an ansatz wave function that provides a zero-momentum state, we can take only quadratic terms $b^{\dagger}_{\bf p}b_{\bf p}$ and neglect the additional boson-boson interaction. Then, the remaining Hamiltonian 
\begin{eqnarray}\label{H}
&&\mathcal{H}=\sum_{|{\bf p}|<\Lambda}\varepsilon_{\bf p}b^{\dagger}_{\bf p}b_{\bf p}-\Delta \sigma_x\nonumber\\
&&+\frac{g_{\uparrow,\Lambda}}{L^d}\sum_{|{\bf p}|,|{\bf p}'|<\Lambda}b^{\dagger}_{\bf p}b_{{\bf p}'}|\uparrow\rangle\langle \uparrow|,
\end{eqnarray}
is formally that for infinite-mass impurity (but $m$ in $\varepsilon_{\bf p}$ is still the reduced boson-impurity mass). Unlike non-interacting fermions, where the spectrum exhibits \cite{Kain_2017, Chen_2025} the mass gap on the Fermi surface, the bosonic counterpart $\varepsilon_{\bf p}$ is smooth. Operator $b^{\dagger}_{\bf p}$ ($b_{\bf p}$) creates (destroys) boson with momentum ${\bf p}$. They satisfy standard commutation relations $[b_{\bf p}, b^{\dagger}_{{\bf p}'}]=\delta_{{\bf p},{\bf p}'}$ and $[b_{\bf p}, b_{{\bf p}'}]=0$. Without the last term, the ground state of (\ref{H}) is the product of Bose-Einstein-condensate state and eigenstate of $\sigma_x$ operator with eigenvalue $+1$: $|\Psi(0)\rangle=\frac{(b^{\dagger}_{\bf 0})^{N}}{\sqrt{N!}}|\textrm{vac}\rangle|+\rangle$ [here $|\textrm{vac}\rangle$ is normalized vacuum for bosons and $|+\rangle=\frac{1}{\sqrt{2}}(|\uparrow\rangle+|\downarrow\rangle)$ stands for spin state]. Therefore, it seems natural to choose this state as the initial one for our quench protocol. Regardless of the seeming simplicity of the considered Hamiltonian (\ref{H}), we had no success in the exact solution of this model. To demonstrate the issues generated by the implication of spin and bosonic degrees of freedom, it is instructive to turn to the Heisenberg representation $\mathcal{O}(t)=e^{it\mathcal{H}}\mathcal{O}e^{-it\mathcal{H}}$. The equations of motion for spin necessarily involve bosonic creation and annihilation operators in the Heisenberg representation. Applying a variant of the Baker-Campbell-Hausdorff formula $e^{it\mathcal{H}}\mathcal{O}e^{-it\mathcal{H}}=\mathcal{O}+it[\mathcal{H},\mathcal{O}]+\frac{(it)^2}{2!}[\mathcal{H},[\mathcal{H},\mathcal{O}]]+\dots$ for $b_{\bf p}(t)$, it is already seen at the level of third term that the transverse magnetic field generates two-body interaction between bosons. Next terms in this series produce all higher-order interactions. In what follows, the time-dependent annihilation operator cannot be written as a linear combination of $b_{\bf p}$s, i.e. $b_{\bf p}(t)\neq \sum_{{\bf p}'}U_{{\bf p},{\bf p}'}(t)b_{{\bf p}'}$.

Unable to solve the problem of spin-boson-bath dynamics exactly, we are forced to make simplifications. Most naturally, the adopted approximation can be explained in Schr\"odinger picture. Starting from the ground state of the non-interacting system, one can obtain the wave function at an arbitrary time moment after the interaction was switched on
\begin{eqnarray}\label{Psi_t}
|\Psi(t)\rangle=e^{-it\mathcal{H}} |\Psi(0)\rangle= e^{it\Delta}\frac{[b^{\dagger}_{\bf 0}(-t)]^{N}}{\sqrt{N!}}|\textrm{vac}\rangle|+\rangle.
\end{eqnarray}
Formally, this is an exact expression, with the approximation scheme demonstrated by calculating various observables. In particular, the wave-function overlap at the initial moment and at an arbitrary time $t$ can be approximated as follows:
\begin{eqnarray}\label{overlap}
&&\langle\Psi(0)|\Psi(t)\rangle=e^{it\Delta} \langle+|\langle\textrm{vac}|\frac{b_{\bf 0}^{N}}{\sqrt{N!}}\frac{[b^{\dagger}_{\bf 0}(-t)]^{N}}{\sqrt{N!}}|\textrm{vac}\rangle |+\rangle\nonumber\\
&&\approx e^{it\Delta} \langle+|[\langle\textrm{vac}|b_{\bf 0}b^{\dagger}_{\bf 0}(-t)|\textrm{vac}\rangle ]^N|+\rangle.
\end{eqnarray}
This is still an equality in a trivial case of no boson-impurity interaction and in the single-boson limit $N=1$. The latter fact guarantees that the single-boson-like approximation (\ref{overlap}) correctly reproduces the linear-in-density term of the exact overlap. Even in the many-boson limit without a transverse magnetic field, our approximation becomes exact. The approximation amounts to replacing the many-body evolution by $N$ copies of the same single-boson orbital, while retaining the full spin dynamics of that orbital. Note that approximation preserves unitarity, i.e. the approximately calculated overlap $\langle\Psi(t)|\Psi(t)\rangle\approx \langle+|[\langle\textrm{vac}|b_{\bf 0}(-t)b^{\dagger}_{\bf 0}(-t)|\textrm{vac}\rangle ]^N|+\rangle$ exactly equals unity. The idea behind this simplification is based on the system's Bose statistics and can be justified by variational calculations. Indeed, even in a case of weakly-interacting bosons with an impurity immersed, the ground state can be well approximated by the mean-field ansatz \cite{Gross_1962, Hryhorchak_2020, Hryhorchak_2020_2, Massignan_2021, Guenther_2021, Schmidt_2022}, where all bosons are assumed to occupy the same elementary orbital, which is computed self-consistently. This is found to be a particularly powerful approach for one-dimensional systems \cite{Volosniev_2017, Pastukhov_2019, Panochko_2019, Petkovic_2022, Petkovic_2023}, where the effects of quantum fluctuations \cite{Jager_2020} are small. The trial wave function that makes the last equality in the overlap (\ref{overlap}) exact is the BEC state $|\Psi(t)\rangle=\frac{\left[B^{\dagger}(t)\right]^N}{\sqrt{N!}}|\textrm{vac}\rangle|+\rangle$, where $B^{\dagger}(t)=\sum_{{\bf p}}U_{{\bf p}}(t)b^{\dagger}_{{\bf p}}$ is the creation operator of boson in the lowest self-consistent orbital at moment $t$. The time-dependent coefficients $U_{{\bf p}}(t)$ (recall $2\times 2$ matrices) are subject to the variational principle \cite{Moccia_1973} with the unitarity constraint $\sum_{{\bf p}}U^{\dagger}_{{\bf p}}(t)U_{{\bf p}}(t)=1$ at equal time arguments.

To discuss the spin dynamics after the interaction quench, we need to calculate the expectation values of the spin components in state $|\Psi(t)\rangle$ [see Eq.~(\ref{Psi_t})]. Fortunately, from the Heisenberg equation of motion for the $\sigma_z$, one can obtain average $\sigma_y$ 
\begin{eqnarray}\label{sigma_y}
\partial_t \langle\sigma_z\rangle=-2\Delta \langle\sigma_y\rangle,
\end{eqnarray}
where $\langle\dots \rangle=\langle\Psi(t)|\dots|\Psi(t)\rangle$. The remaining two averages can be approximately computed in the same manner as the overlap (\ref{overlap}). Using the self-reversibility of Pauli matrices $\sigma_{\alpha}^2=1$, we can write
\begin{eqnarray}\label{sigma_alpha}
&&\langle\sigma_{\alpha}\rangle= \langle+|\langle\textrm{vac}|\frac{[b_{\bf 0}(-t)]^{N}}{\sqrt{N!}}\sigma_{\alpha}\frac{[b^{\dagger}_{\bf 0}(-t)]^{N}}{\sqrt{N!}}|\textrm{vac}\rangle |+\rangle\nonumber\\
&&=\langle+|\langle\textrm{vac}|\frac{[b_{\bf 0}(-t)]^{N}}{\sqrt{N!}}\frac{[\sigma_{\alpha}b^{\dagger}_{\bf 0}(-t)\sigma_{\alpha}]^{N}}{\sqrt{N!}}|\textrm{vac}\rangle \sigma_{\alpha}|+\rangle\nonumber\\
&&\approx \langle+|[\langle\textrm{vac}|b_{\bf 0}(-t)\sigma_{\alpha}b^{\dagger}_{\bf 0}(-t)\sigma_{\alpha}|\textrm{vac}\rangle ]^N\sigma_{\alpha}|+\rangle.
\end{eqnarray}
Again, for $N=1$, the last equality is exact. Similarly to the appropriate expression in the overlap, expectation value $\langle\textrm{vac}|b_{\bf 0}(-t)\sigma_{\alpha}b^{\dagger}_{\bf 0}(-t)\sigma_{\alpha}|\textrm{vac}\rangle$ appeals to a single-particle problem and can be calculated exactly. From the above discussion of the single-boson limit, it is clear that the generic structure of these vacuum expectation values is the following $1+(\dots)/L^d$ at large but finite volumes $L^d$. Therefore, in the thermodynamic limit, the overlap (\ref{overlap}) and average spins can be written in exponential form
\begin{subequations}
\begin{align}
    \langle\Psi(0)|\Psi(t)\rangle\approx e^{it\Delta} \langle+|e^{\varphi(t)}|+\rangle, \label{eq:2.15a}\\
    \langle\sigma_{\alpha}\rangle\approx \langle+|e^{\varphi_{\alpha}(t)}\sigma_{\alpha}|+\rangle. \label{eq:2.15b}
\end{align}
\end{subequations}
All introduced $2\times 2$ matrices $\varphi(t)$ and $\varphi_{\alpha}(t)$ are strictly linear in density $n=N/L^d$ of bosons and chosen in a way to satisfy initial conditions $\varphi(0)=\varphi_{\alpha}(0)=0$. Their explicit expressions are cumbersome and constructed from the $T$-matrix elements $\langle {\bf p}|\mathcal{T}(\omega)|{\bf p}'\rangle$ in addition to a combination of $\Pi(\omega\pm\Delta)$ with the subsequent integration over the frequency. Note that our calculations are not restricted to the initial spin state $|+\rangle$, and can be easily extended to an arbitrary point on the Bloch sphere. One should keep in mind that the approximate mean-field-like decomposition of the many-body expectation values is exact \cite{Hryhorchak_2026} for the system without Rabi coupling. Finally, it is also valid in the case of switched-off boson-impurity interaction. Therefore, we expect reliable results, at least in two limits: small $\Delta$s and large two-body couplings, and vice versa.

\section{Results and discussion}
In general, the ground state of an ideal Bose gas is insensitive to the spatial dimension; however, the impurity behavior differs \cite{Panochko_2021, Panochko_2022, Hryhorchak_2023_2} for $d\le 2$ and $d>2$. In particular, low-dimensional systems of non-interacting bosons do not `feel' a static point-like impurity in their ground BEC state. Therefore, the results for the two examples, $d=2$ and $d=3$, are considered separately below.

\subsection{3D system}
Calculations of key observables--the time-dependence of the wave-function overlap and average spin components--rely on the structure of the boson-impurity $T$-matrix in the complex plane and require numerical computations of the frequency integrals. Fortunately, for our purposes, the $\omega$-dependent function that determines the $T$-matrix element can be calculated analytically \cite{Mulkerin_2024} $\tau^{-1}(\omega)=g^{-1}_{\uparrow}\left\{1-\sqrt{-\omega-\Delta}/2\sqrt{|\epsilon_{\uparrow}|}-\sqrt{-\omega+\Delta}/2\sqrt{|\epsilon_{\uparrow}|}\right\}$. Here $g_{\uparrow}=2\pi a_{\uparrow}/m$ is the `observable' coupling with $a_{\uparrow}$ being the boson-impurity $s$-wave scattering length, which coincides with the width of the two-body bound state [with energy $\epsilon_{\uparrow}=-1/2ma^2_{\uparrow}$] without Rabi coupling. Real negative-valued zero of braces in $\tau^{-1}(\omega)$ determines the bound state energy $|\epsilon_{\Delta}|/|\epsilon_{\uparrow}|=1+(\Delta/2\epsilon_{\uparrow})^2$ in the presence of Rabi coupling. Notably, the intense $\Delta>2|\epsilon_{\uparrow}|$ transverse field acting on the spin breaks down the bound state. Above this magnitude of Rabi coupling, the many-body system restores its thermodynamic character even at equilibrium. Numerical procedure suggests the diagonalization of matrices 
$\varphi(t)$ and $\varphi_{\alpha}(t)$ at an arbitrary time moment, and then calculation of their exponentiated expectation values in state $|+\rangle$.

\begin{figure}[h!]
	\includegraphics[width=0.235\textwidth]{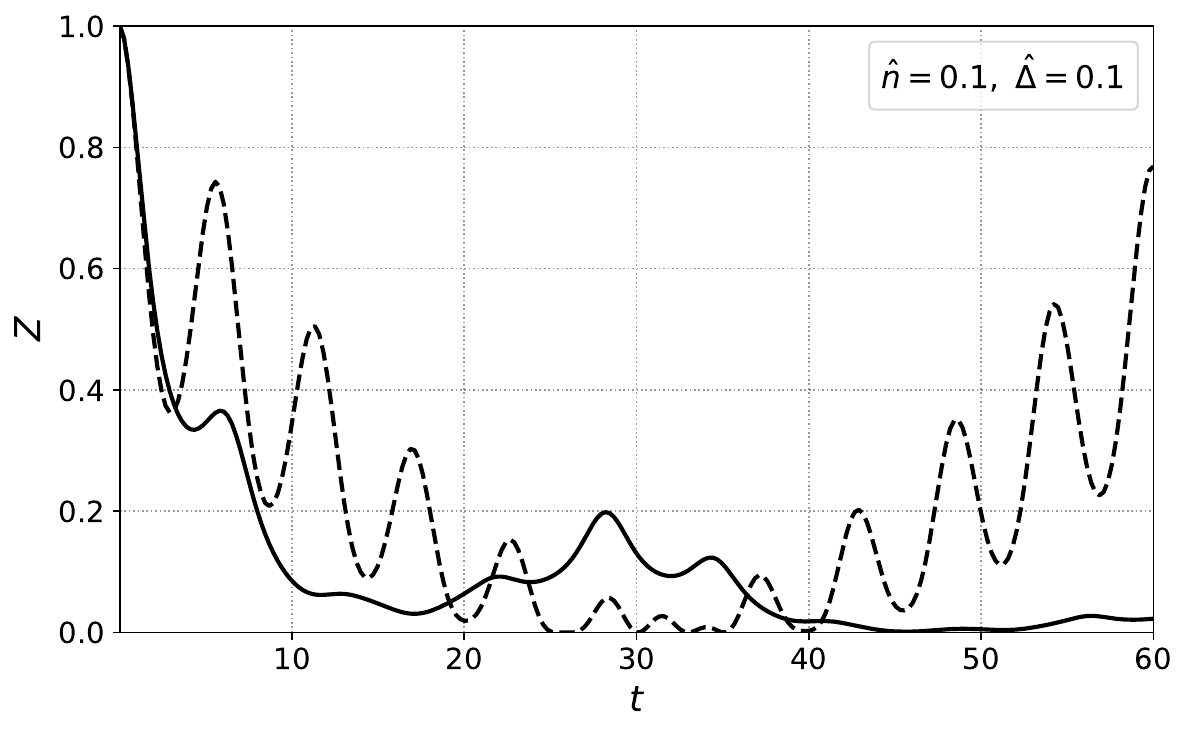}
    \includegraphics[width=0.235\textwidth]{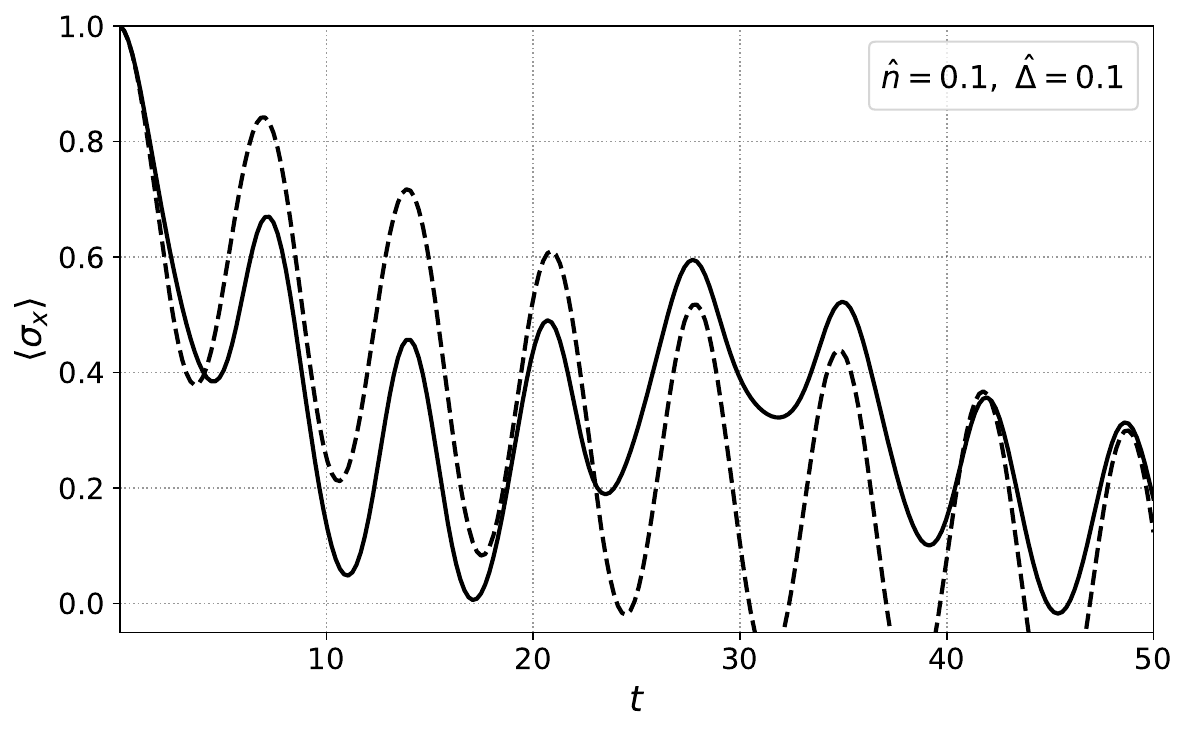}
    \includegraphics[width=0.235\textwidth]{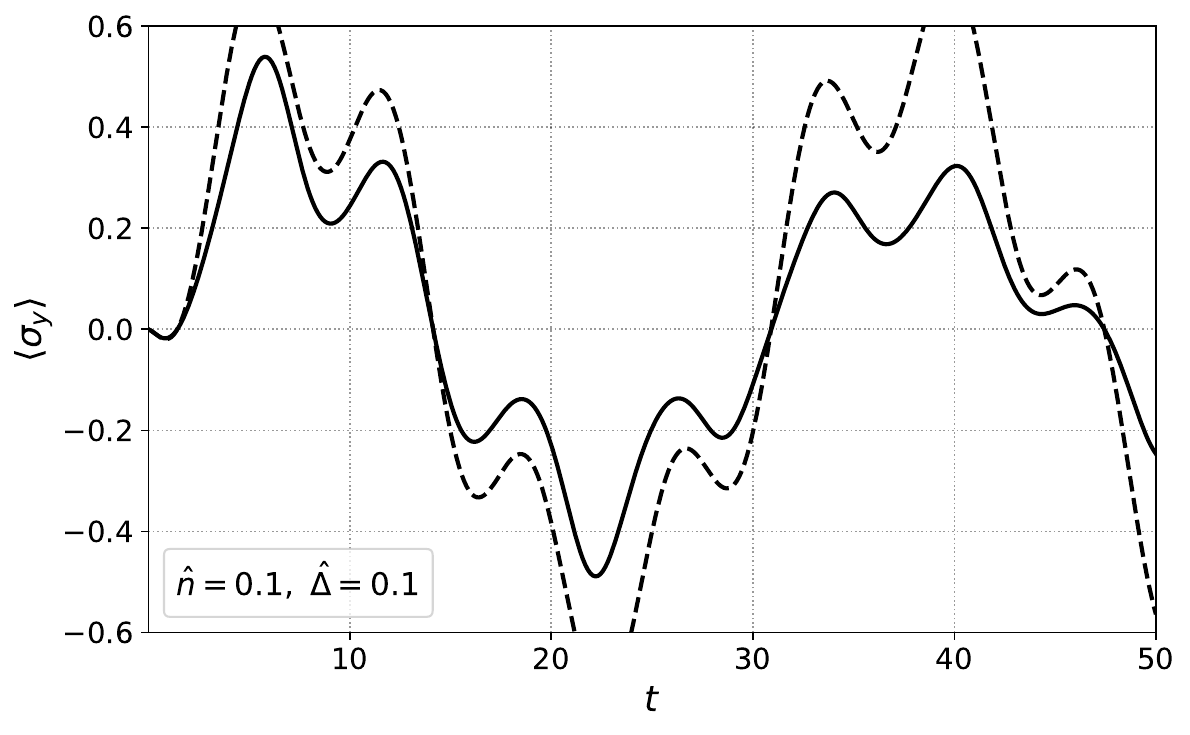}
    \includegraphics[width=0.235\textwidth]{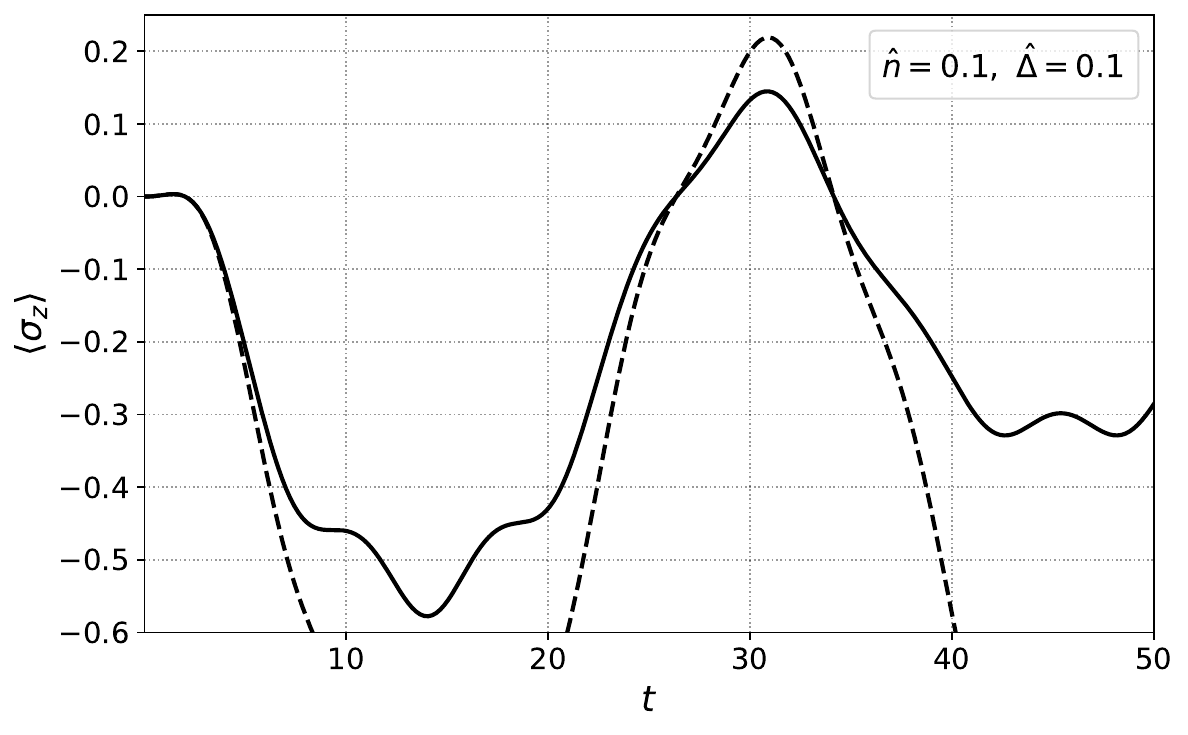}
	\caption{Time dependence of the modulus squared wave-function overlap $Z(t)$ and average spin components $\langle\sigma_{\alpha}\rangle$ for set of parameters $\hat{n}=0.1$ and $\hat{\Delta}=0.1$ (see text). Solid lines correspond to our mean-field-like approach, while dashed lines represent the linear in density of bosons expansion of the exact result.}
	\label{fig:3D_n=0.1_D=0.1}
\end{figure}

\begin{figure}[h!]
	\includegraphics[width=0.235\textwidth]{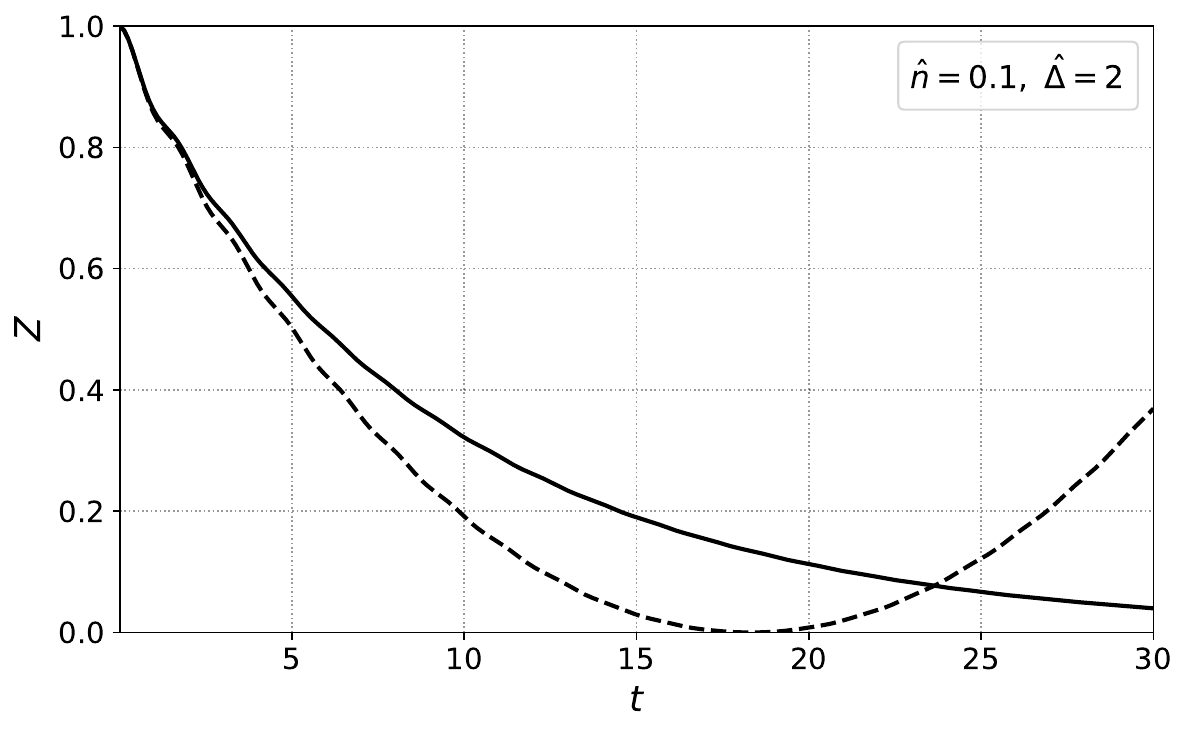}
    \includegraphics[width=0.235\textwidth]{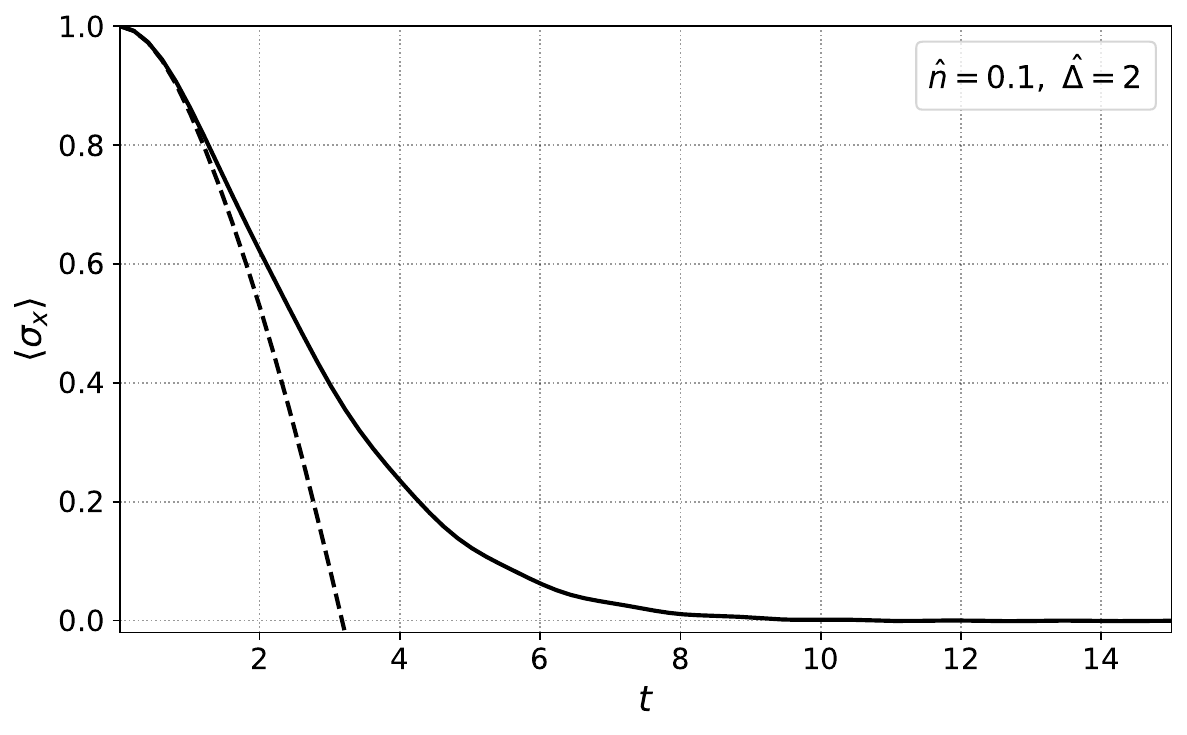}
    \includegraphics[width=0.235\textwidth]{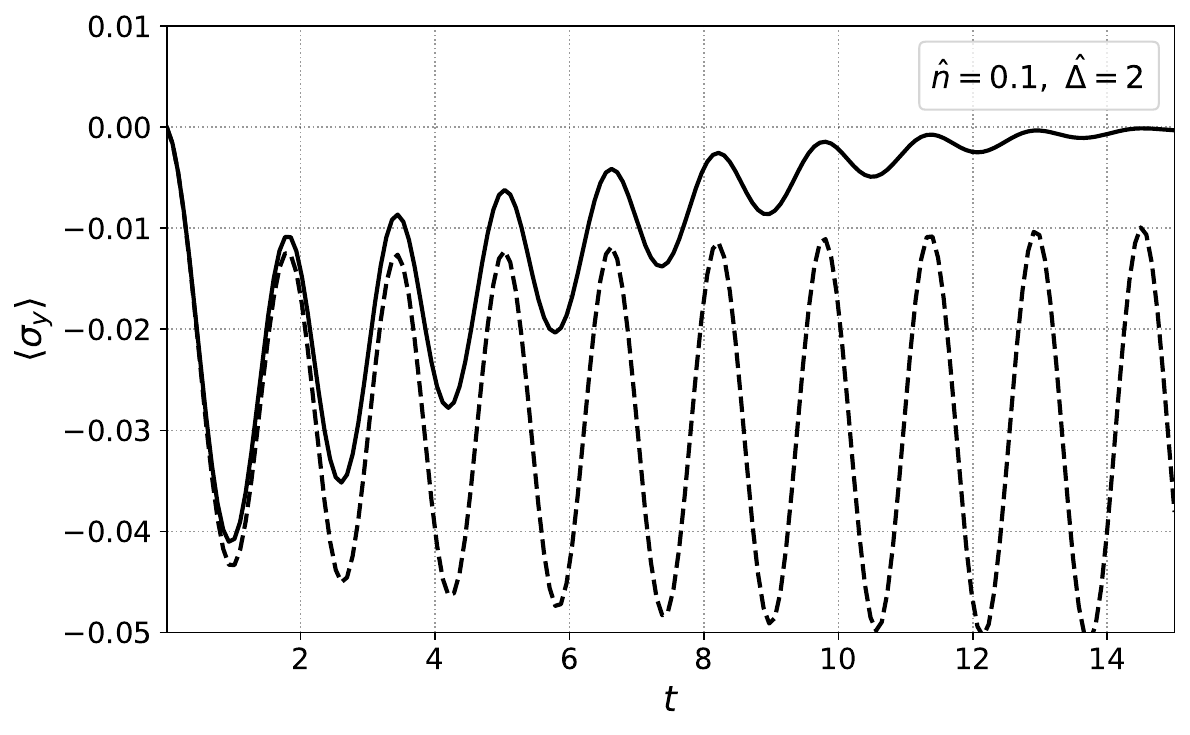}
    \includegraphics[width=0.235\textwidth]{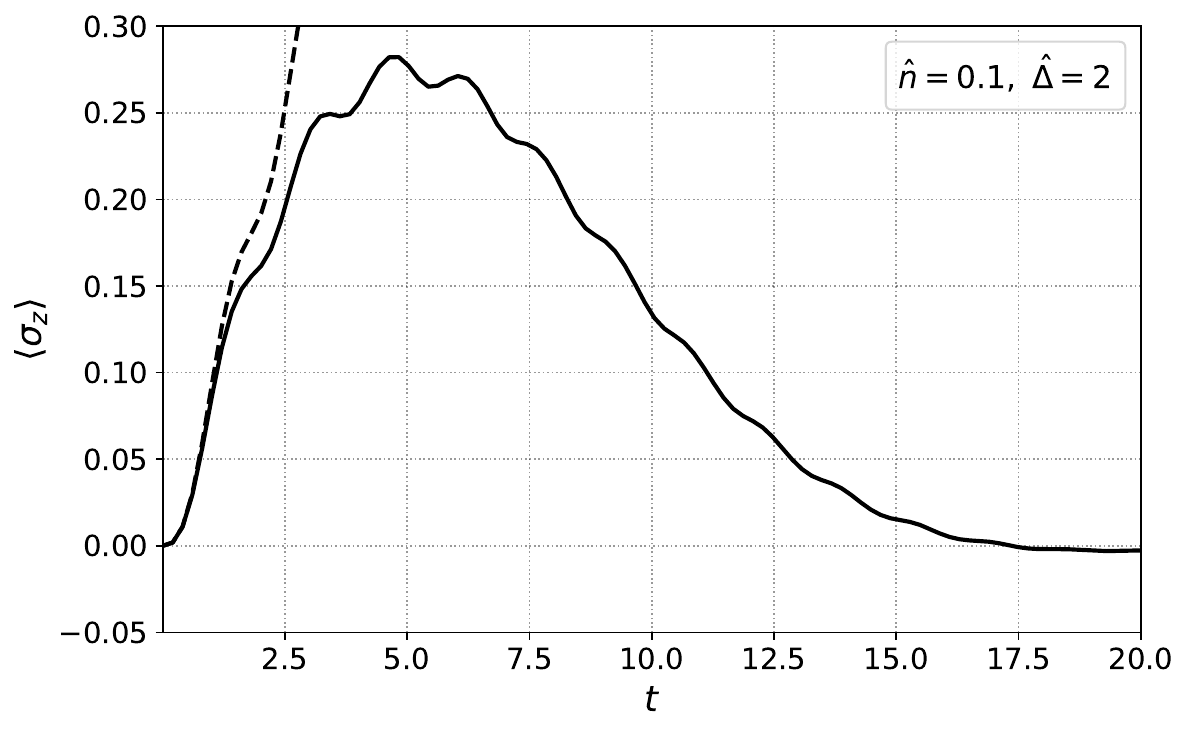}
	\caption{Same as in Fig.~\ref{fig:3D_n=0.1_D=0.1} but for another set of parameters $\hat{n}=0.1$ and $\hat{\Delta}=2$.}
	\label{fig:3D_n=0.1_D=2}
\end{figure}

\begin{figure}[h!]
	\includegraphics[width=0.235\textwidth]{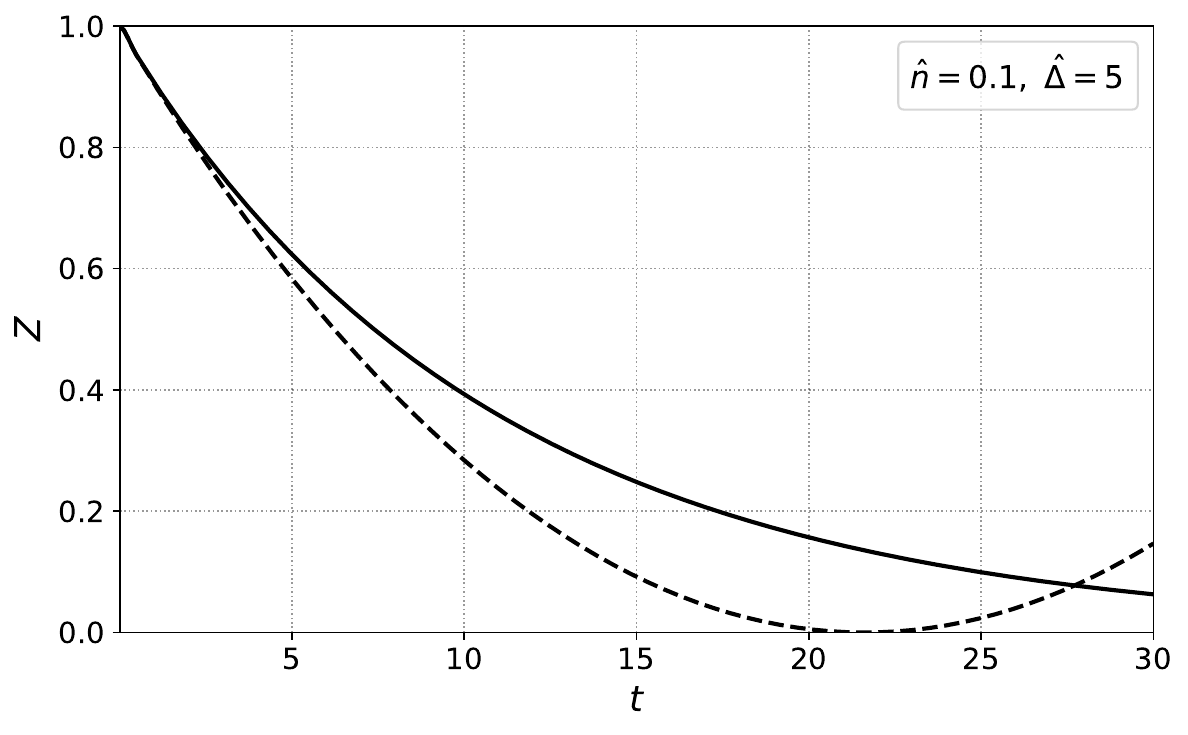}
    \includegraphics[width=0.235\textwidth]{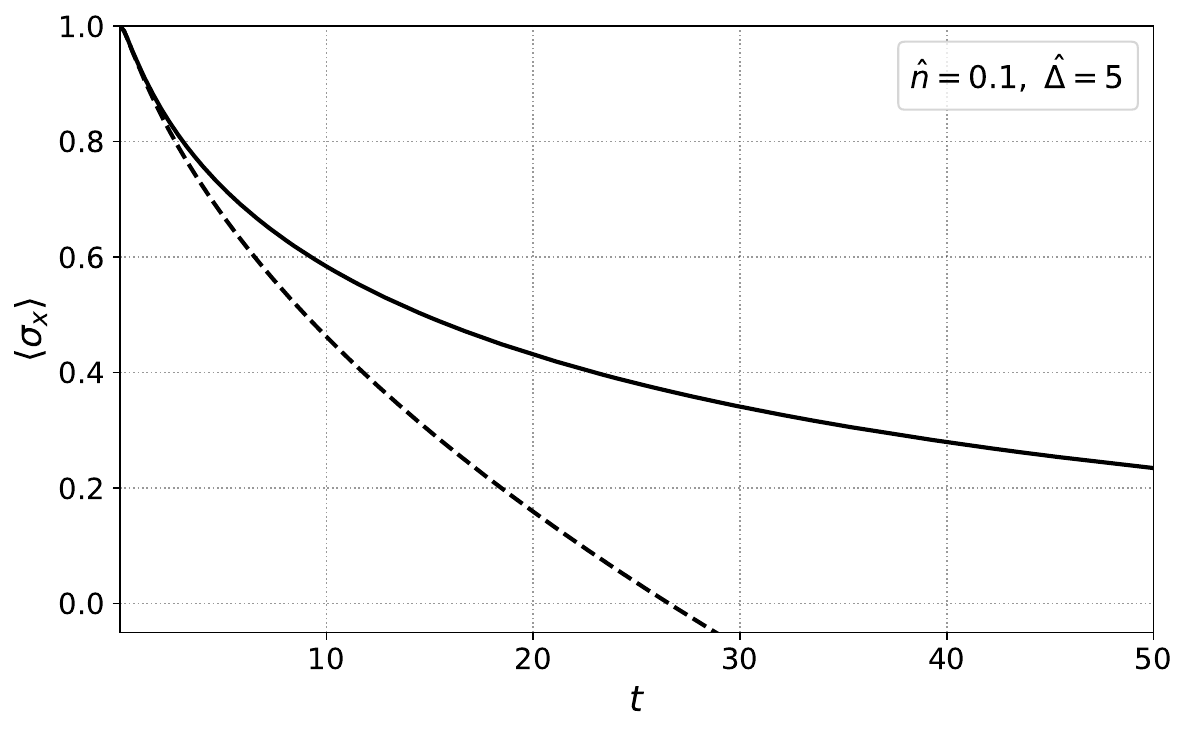}
    \includegraphics[width=0.235\textwidth]{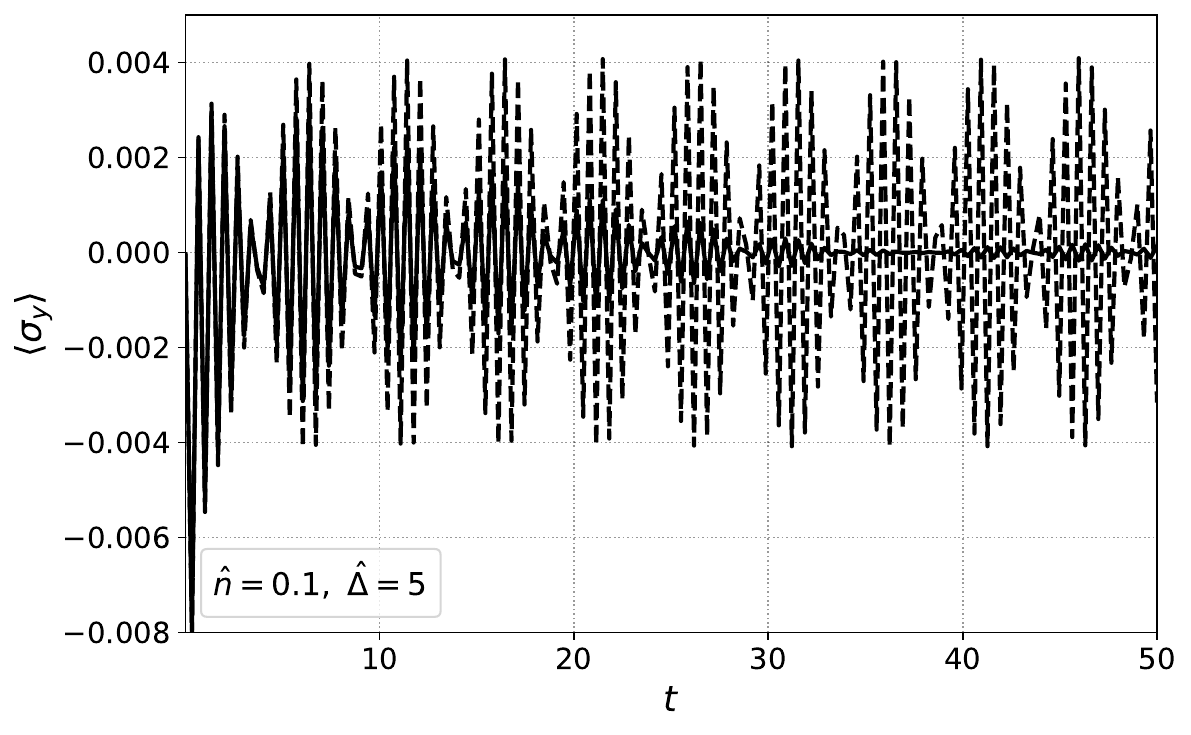}
    \includegraphics[width=0.235\textwidth]{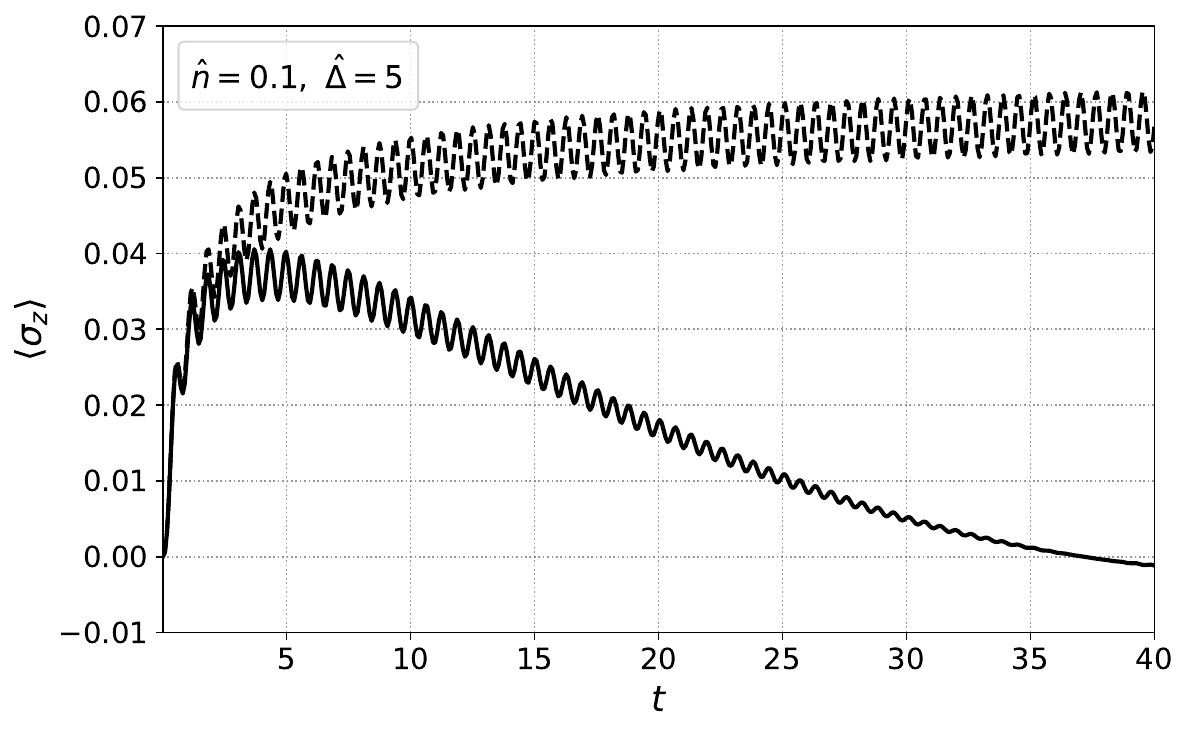}
	\caption{Same density of the system as in Fig.~\ref{fig:3D_n=0.1_D=0.1} but at larger Rabi coupling $\hat{\Delta}=5$.}
	\label{fig:3D_n=0.1_D=5}
\end{figure}

\begin{figure}[h!]
	\includegraphics[width=0.235\textwidth]{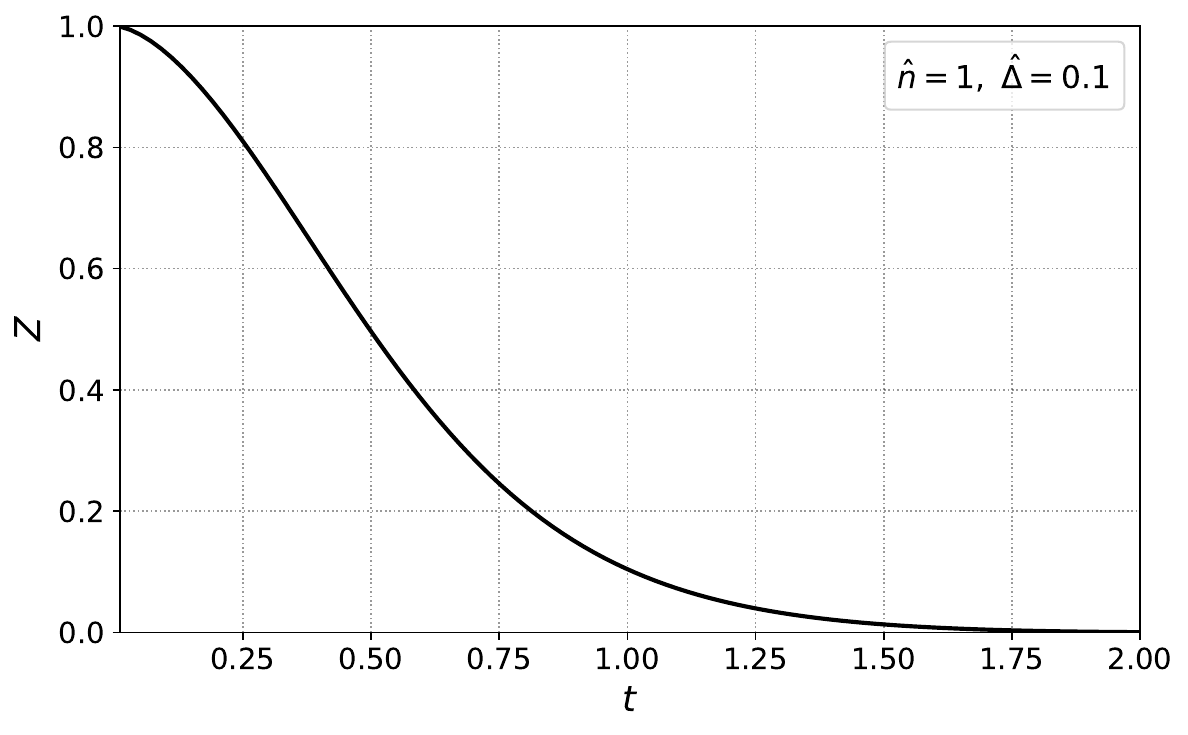}
    \includegraphics[width=0.235\textwidth]{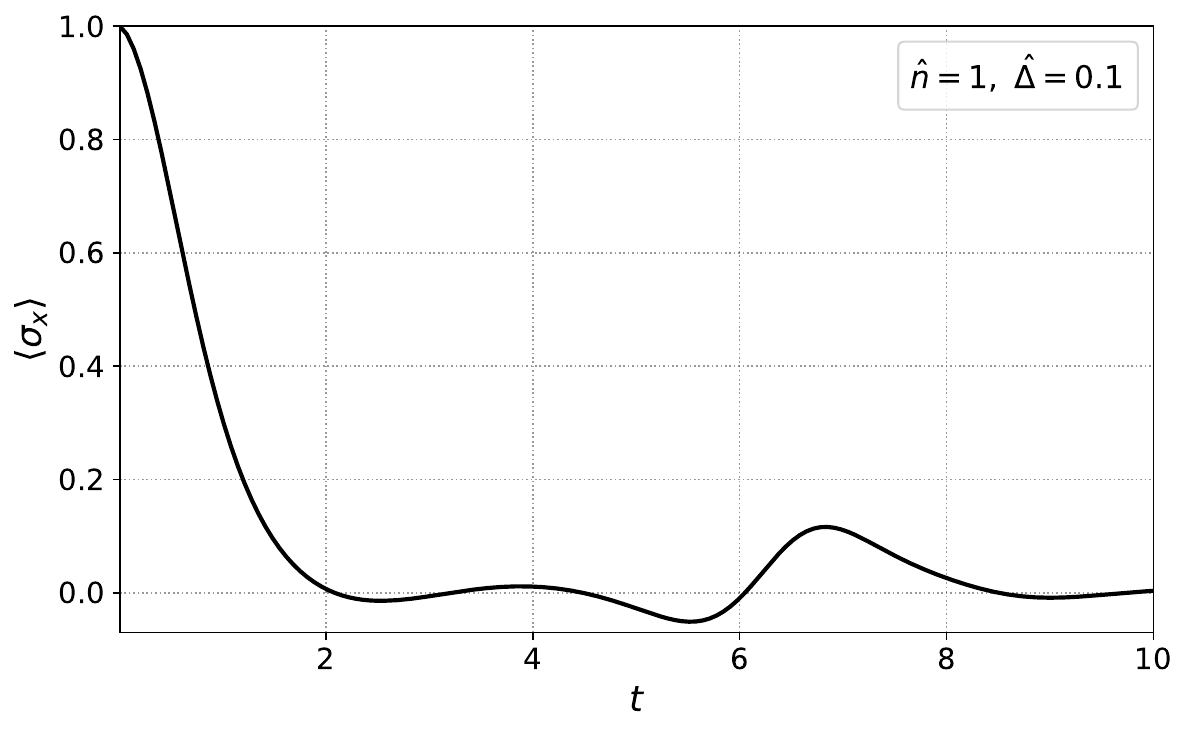}
    \includegraphics[width=0.235\textwidth]{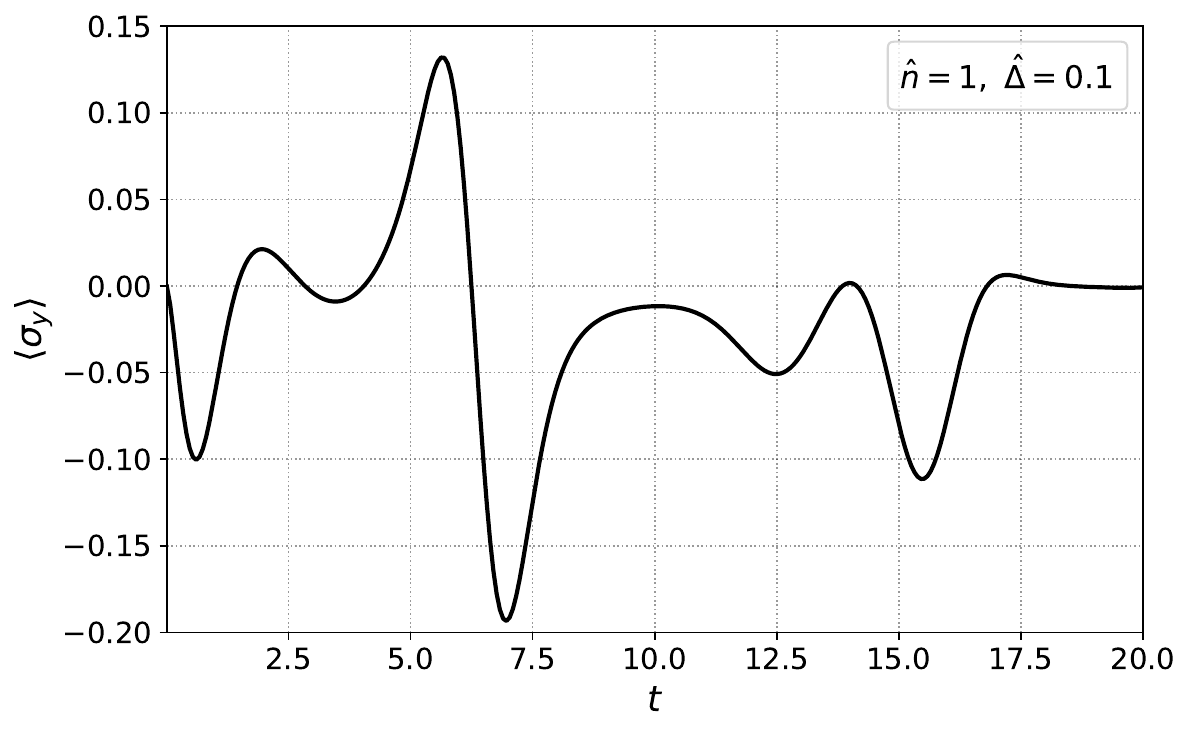}
    \includegraphics[width=0.235\textwidth]{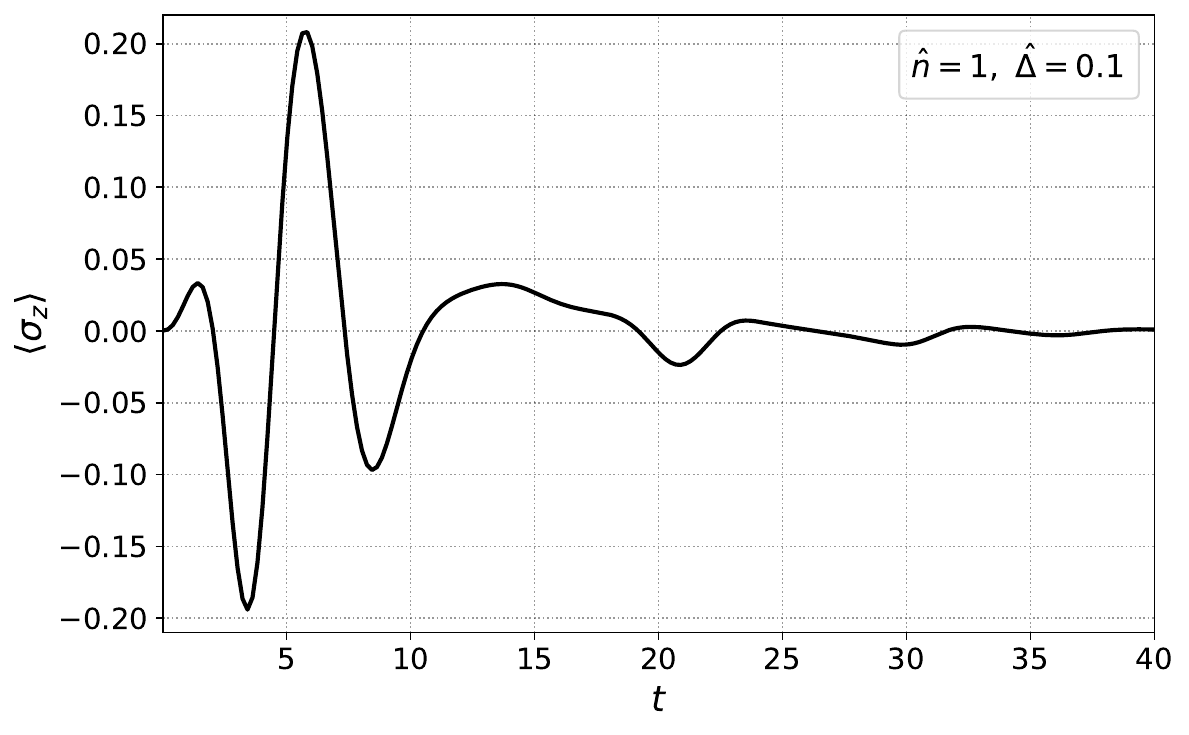}
	\caption{Impurity observables for dimensionless coupling $\hat{n}=1$ and Rabi field $\hat{\Delta}=0.1$.}
	\label{fig:3D_n=1_D=0.1}
\end{figure}

\begin{figure}[h!]
	\includegraphics[width=0.235\textwidth]{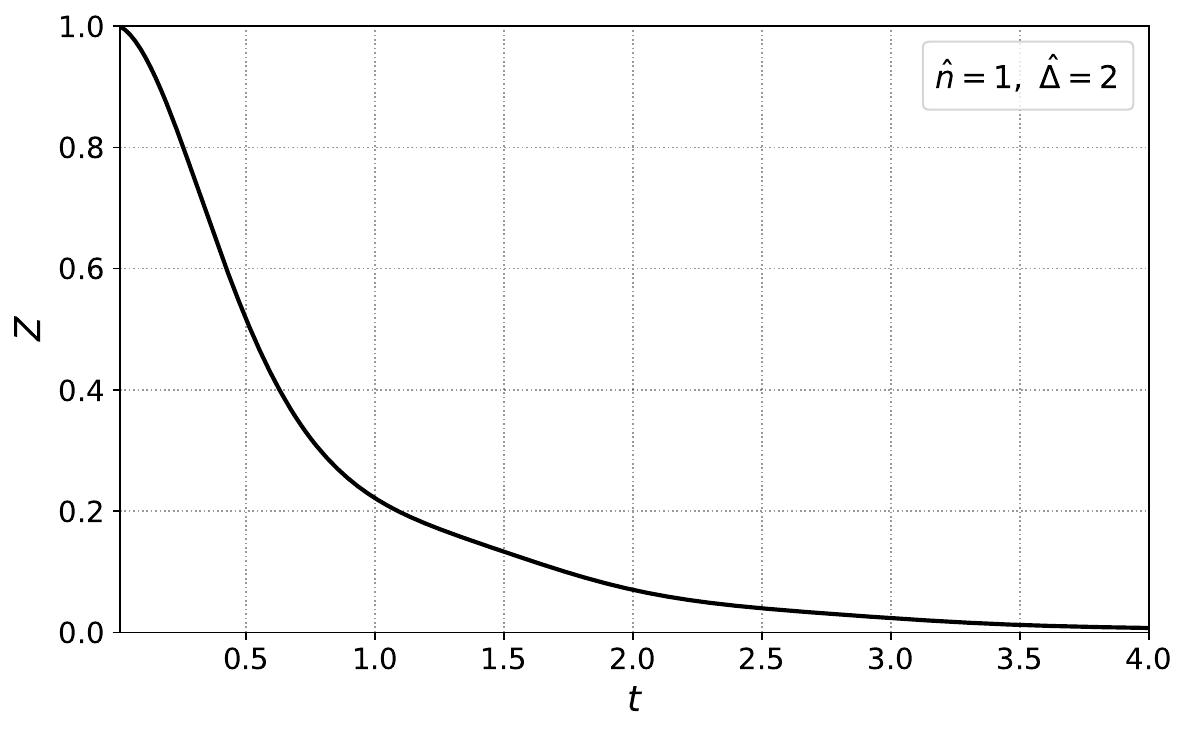}
    \includegraphics[width=0.235\textwidth]{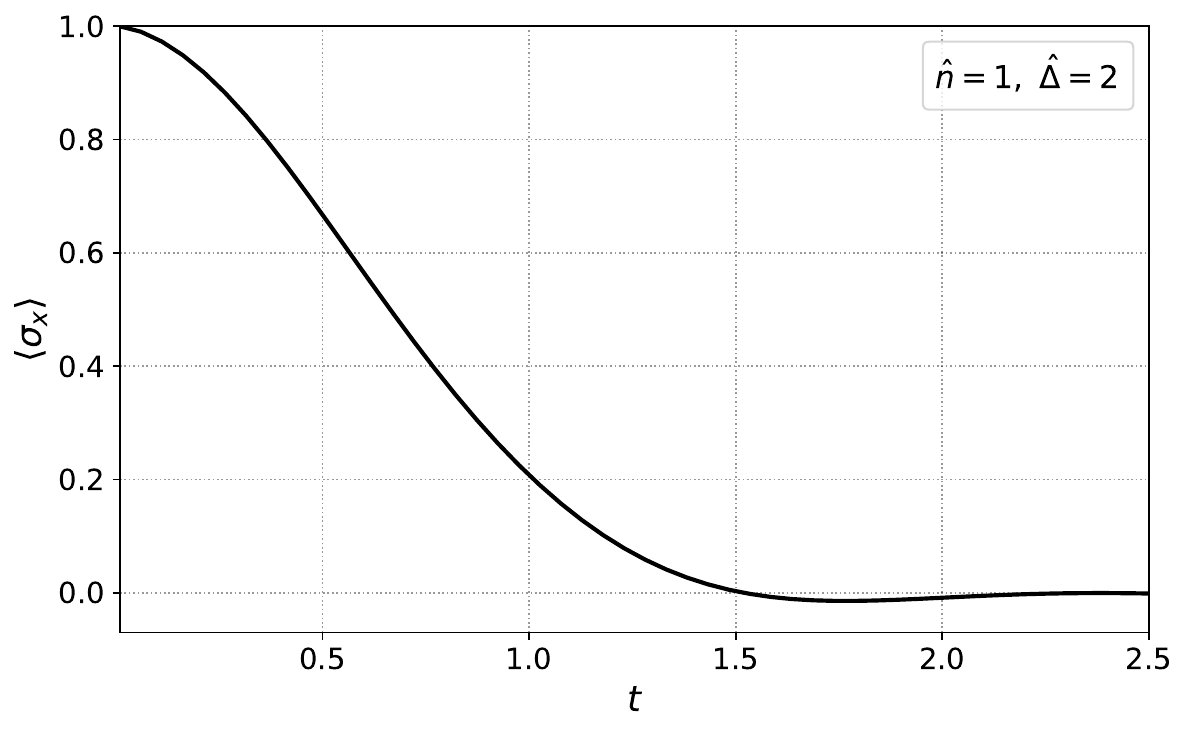}
    \includegraphics[width=0.235\textwidth]{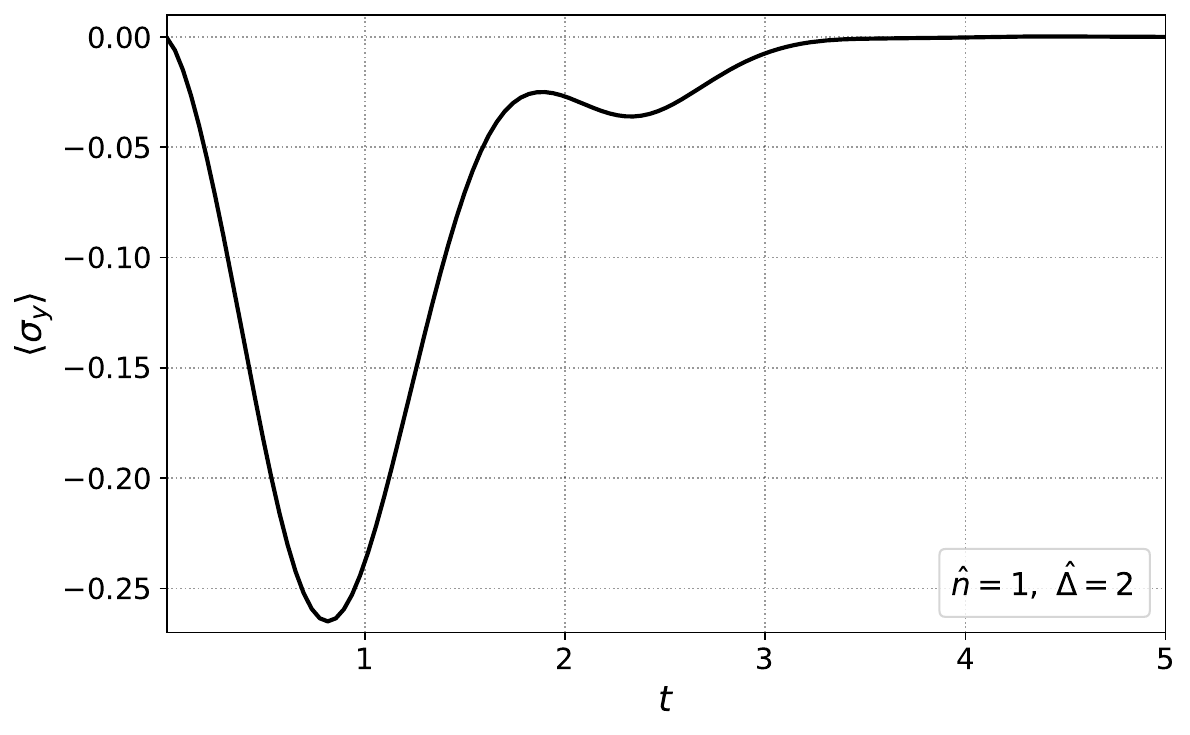}
    \includegraphics[width=0.235\textwidth]{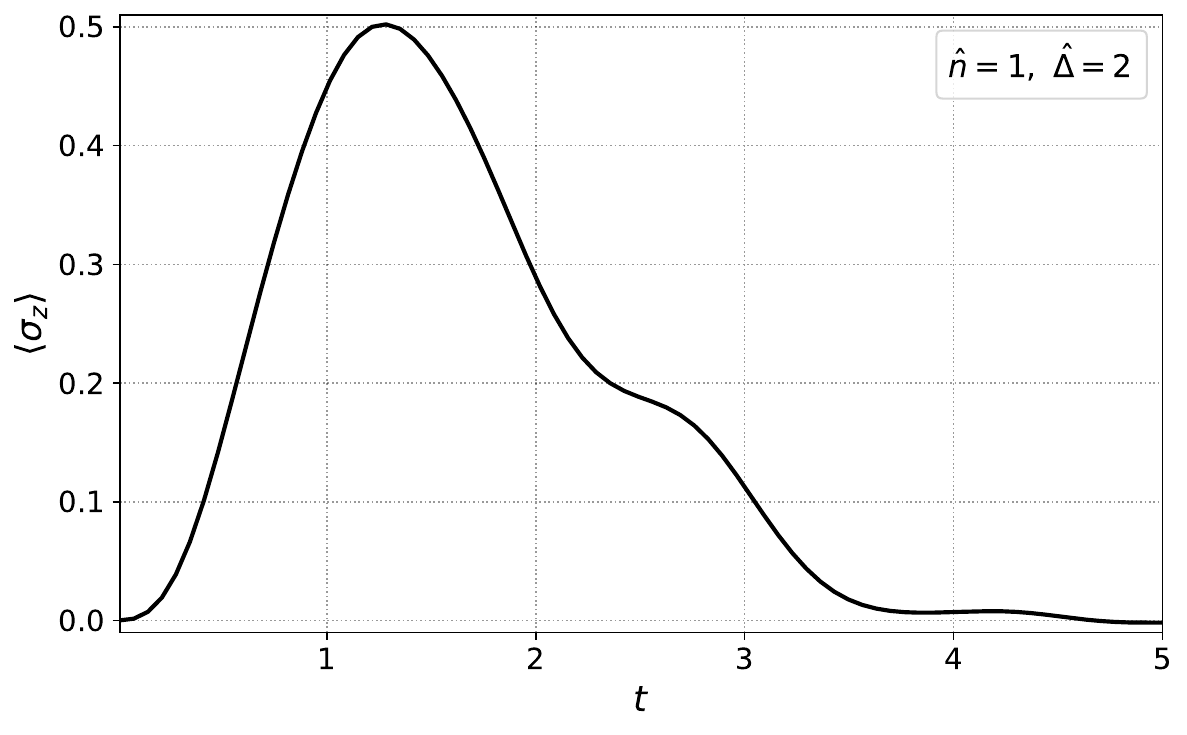}
	\caption{Same density of the system as in Fig.~\ref{fig:3D_n=1_D=0.1} and resonant Rabi coupling $\hat{\Delta}=2$.}
	\label{fig:3D_n=1_D=2}
\end{figure}

\begin{figure}[h!]
	\includegraphics[width=0.235\textwidth]{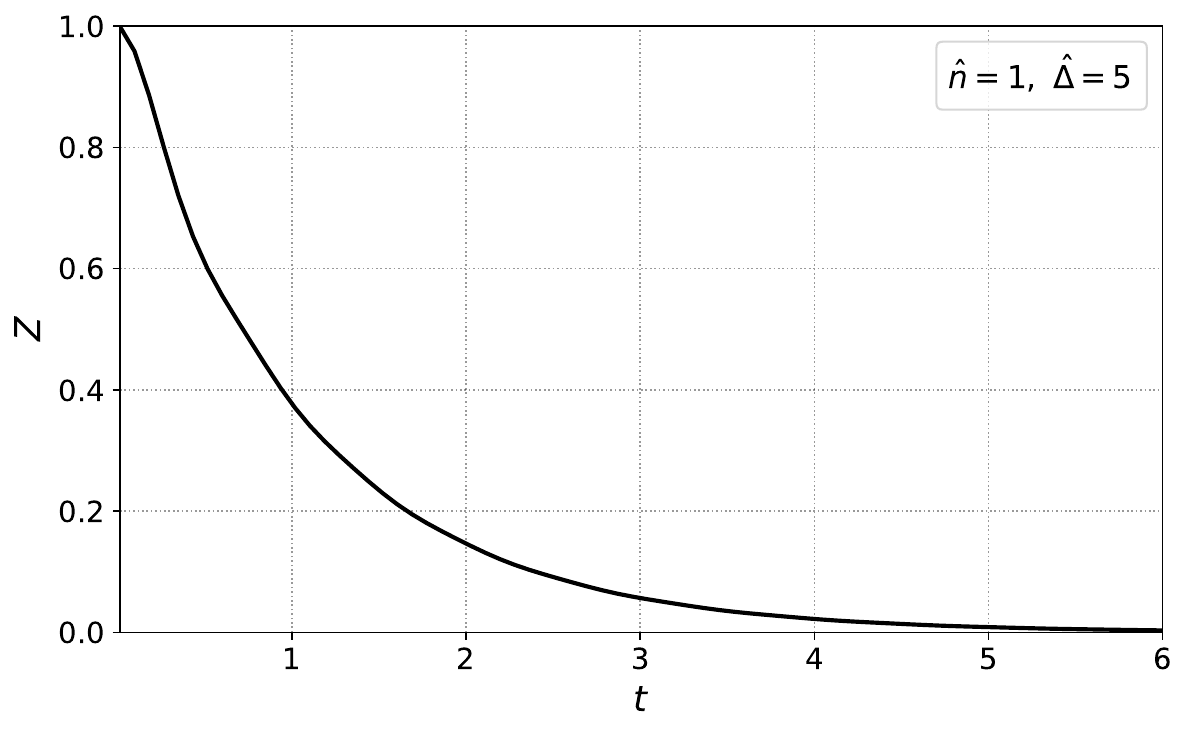}
    \includegraphics[width=0.235\textwidth]{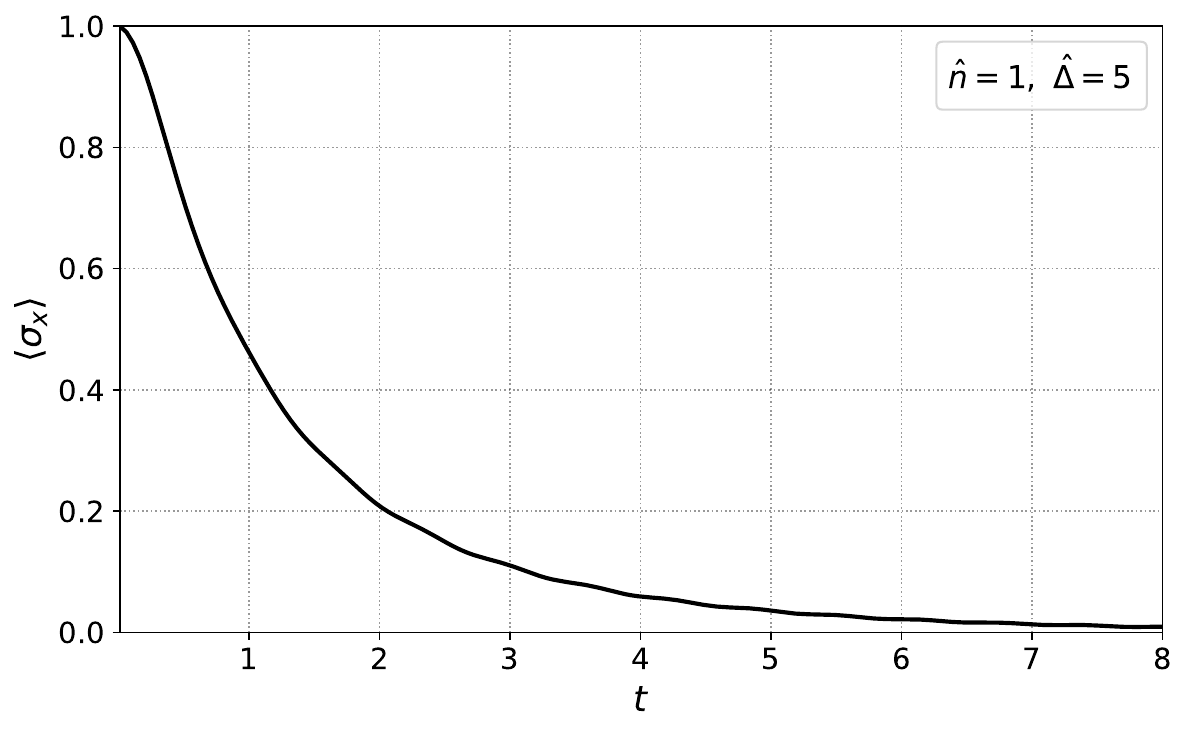}
    \includegraphics[width=0.235\textwidth]{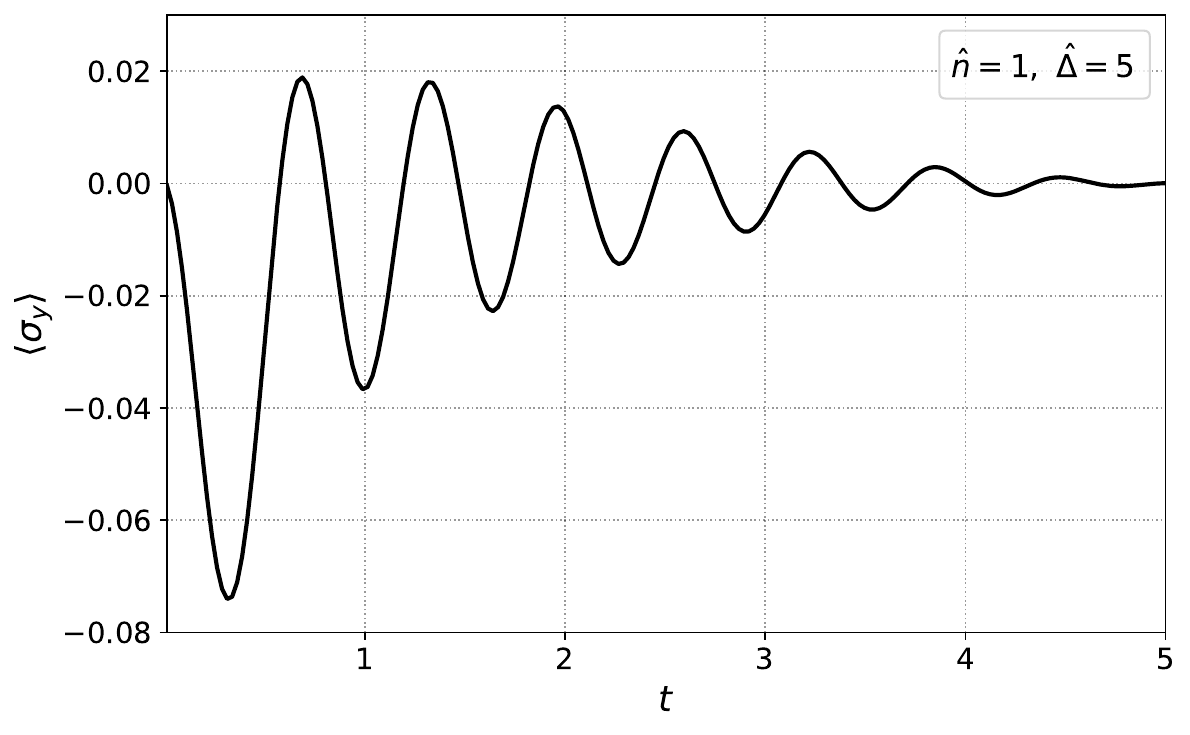}
    \includegraphics[width=0.235\textwidth]{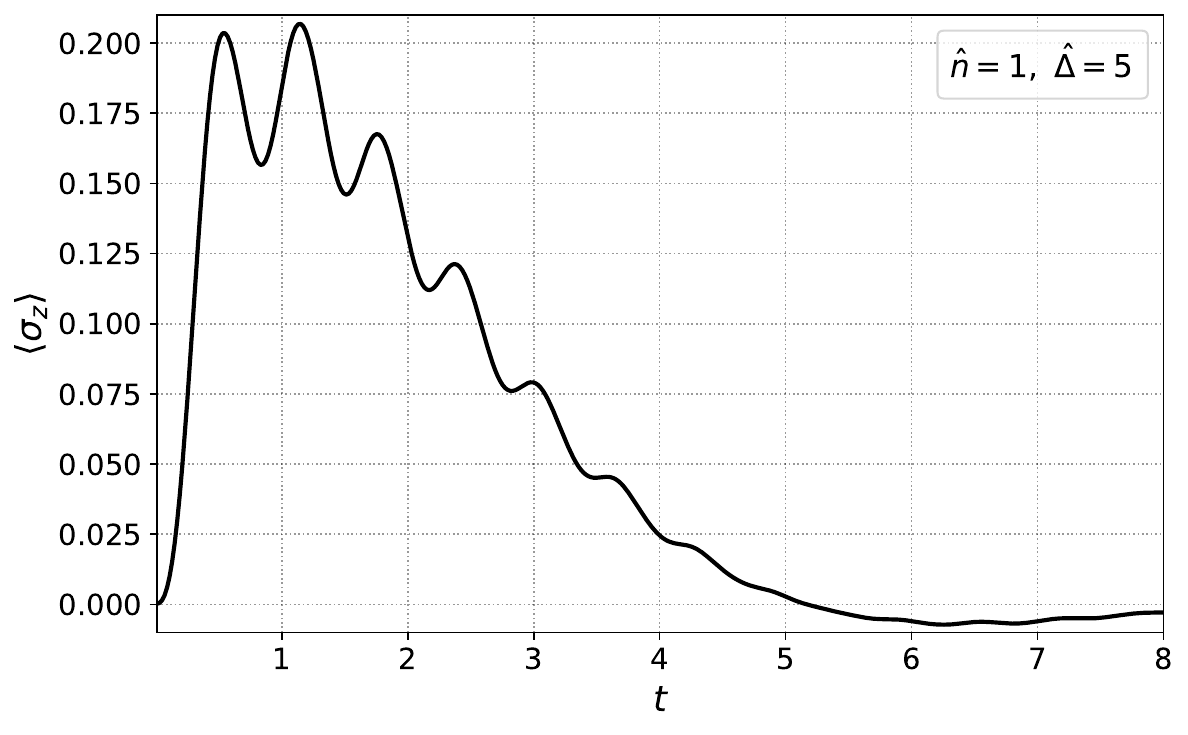}
	\caption{Same density of the system as in Fig.~\ref{fig:3D_n=1_D=0.1} but at larger transverse field $\hat{\Delta}=5$.}
	\label{fig:3D_n=1_D=5}
\end{figure}

For numerical calculations, we introduced the dimensionless density of bosons $\hat{n}=g_{\uparrow}n/|\epsilon_{\uparrow}|$, which measures the intensity of boson-impurity coupling, and the strength of Rabi coupling $\hat{\Delta}=\Delta/|\epsilon_{\uparrow}|$. Figure~\ref{fig:3D_n=0.1_D=0.1}
demonstrates the behavior of the squared modulus of the wave-function overlap $Z(t)=|\langle\Psi(0)|\Psi(t)\rangle|^2$ and expectation values of the impurity spin components at different time moments (in units of $|\epsilon_{\uparrow}|^{-1}$) for a relatively dilute system $\hat{n}=0.1$ and weak Rabi field $\hat{\Delta}=0.1$. For comparison, we also presented the leading-order result (dashed curves), which is linear in the boson density and becomes exact in the limit of extreme boson dilution. In general, we see that the linear approximation is inadequate at late times despite weak boson-impurity coupling and smallness of the Rabi field. This discrepancy is even more visible (see Fig.~\ref{fig:3D_n=0.1_D=2}) at the unitary limit $\hat{\Delta}=2$, where the width of the boson-impurity bound state diverges.
Without the bound state formed, the system displays a faster trend towards thermalization. For even larger Rabi couplings--the limit where our mean-field calculations become exact--one observes (see Fig.~\ref{fig:3D_n=0.1_D=5}) the tendency toward steady state formation. 
This behavior is in strict contrast to the linear-over-density approximation. It should be noted that, unlike the inclusion of only a linear-in-density term, our approach gives reasonable results at any boson-impurity couplings. To study the effects of higher boson densities, we performed extensive calculations of all the above observables at $\hat{n}=1$. Figures~\ref{fig:3D_n=1_D=0.1}, \ref{fig:3D_n=1_D=2}, \ref{fig:3D_n=1_D=5}
summarize the outcome of numerical computations for several dimensionless Rabi couplings (here, the linear in density approximation works extremely badly except at the initial times, therefore not presented). Again, for a large imbalance between the two parameters $\hat{n}$ and $\hat{\Delta}$, one should expect a reliable match between the obtained approximate time dependencies and the exact system dynamics. Therefore, two examples, $\hat{\Delta}=0.1$ and $\hat{\Delta}=5$ (see Fig.~\ref{fig:3D_n=1_D=0.1} and Fig.~\ref{fig:3D_n=1_D=5} appropriately), should be of particular interest for the prediction of the spin behavior. In three cases, we observe spin decoherence dynamics due to interaction with bosons and, consequently, hints of the steady-state formation. Comparing cases with the well-defined boson-impurity bound state $\hat{\Delta}=0.1$, resonant (unitary) limit $\hat{\Delta}=2$, and system with only continuous spectrum $\hat{\Delta}=5$ in the bosonic sector, we conclude that the bound-state presence generally postpones the spin thermalization. Large oscillations in the $\langle\sigma_y\rangle$ and $\langle\sigma_z\rangle$ behaviors presented in Fig.~\ref{fig:3D_n=1_D=5} (and their less noticeable fingerprints in Fig.~\ref{fig:3D_n=1_D=2}) are intrinsic for an isolated spin dynamics prepared in the initial state $|+\rangle$ and evolved by a transverse magnetic field; the interaction with bosons leads to an exponential decay of their amplitudes.

\subsection{2D system}
Similarly to the three-dimensional case, the 2D two-body $T$-matrix of a spinless atom and a spinful impurity can be obtained analytically $\tau^{-1}(\omega)=\frac{m}{2\pi}\ln \frac{|\epsilon_{\uparrow}|}{\sqrt{(-\omega)^2-\Delta^2}}$. In this geometry, a single spherically-symmetric bound state with energy $|\epsilon_{\Delta}|/|\epsilon_{\uparrow}|=\sqrt{1+(\Delta/\epsilon_{\uparrow})^2}$ is robust to any strength of Rabi coupling predetermining the system's long-time evolution.

\begin{figure}[h!]
	\includegraphics[width=0.235\textwidth]{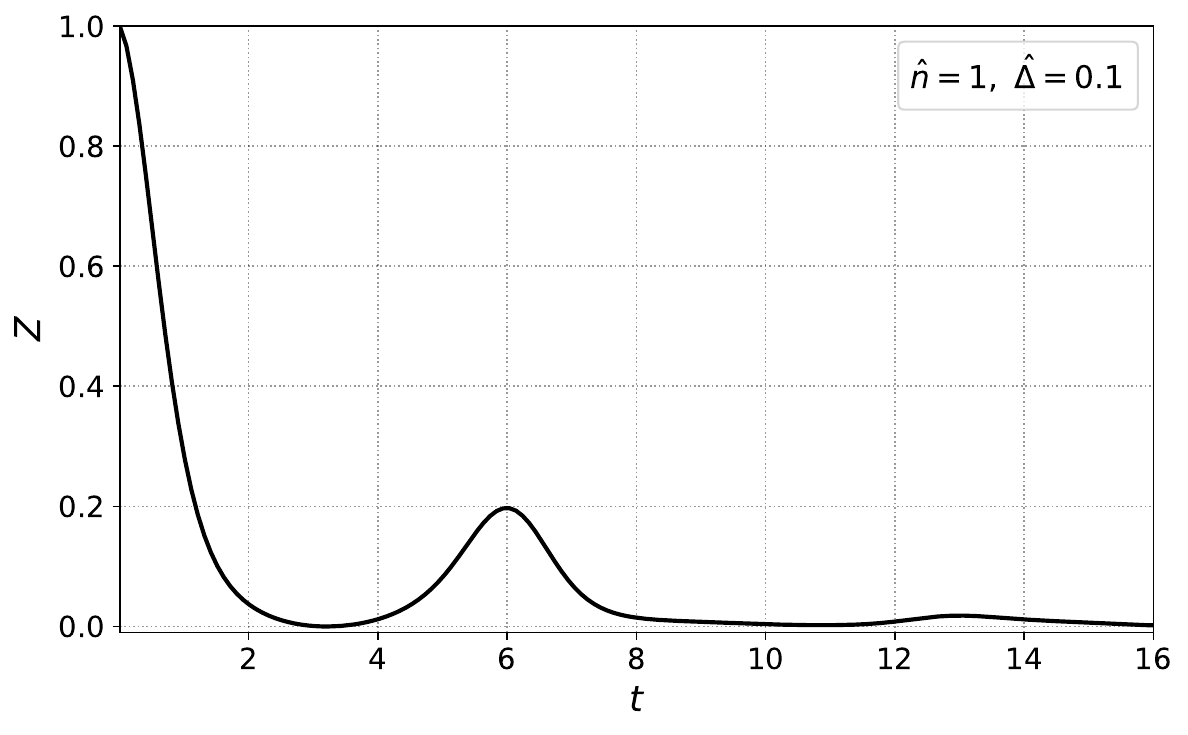}
    \includegraphics[width=0.235\textwidth]{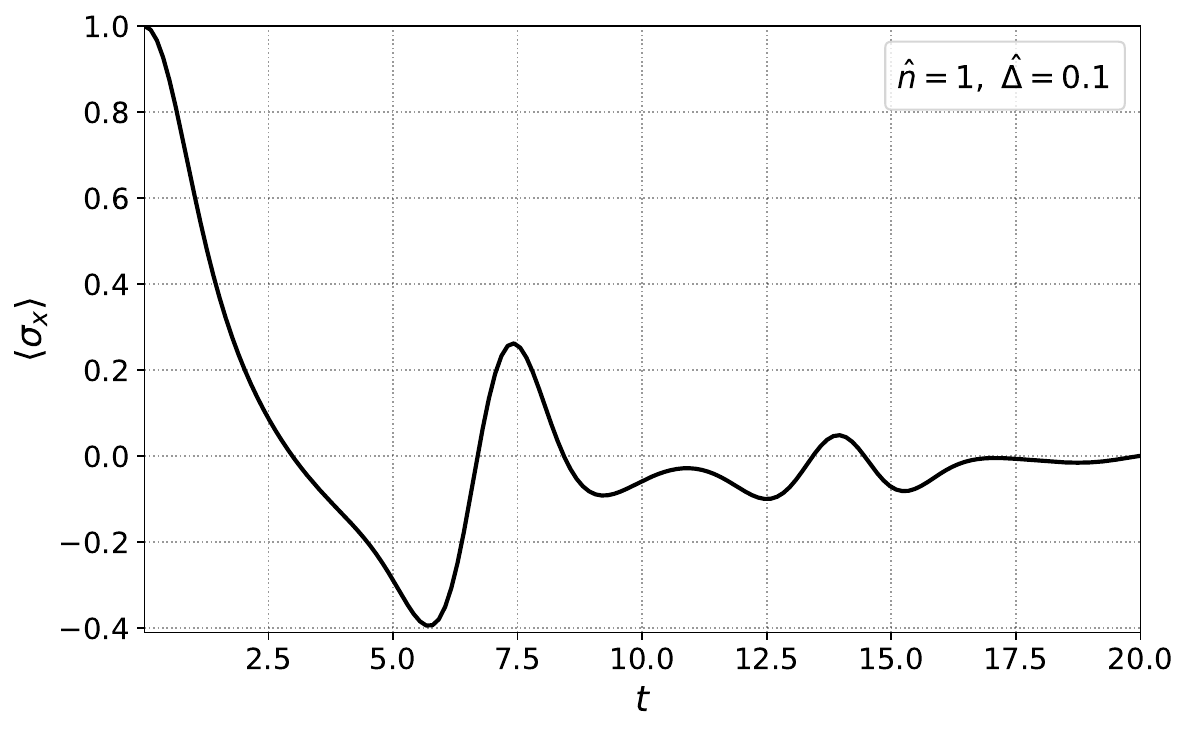}
    \includegraphics[width=0.235\textwidth]{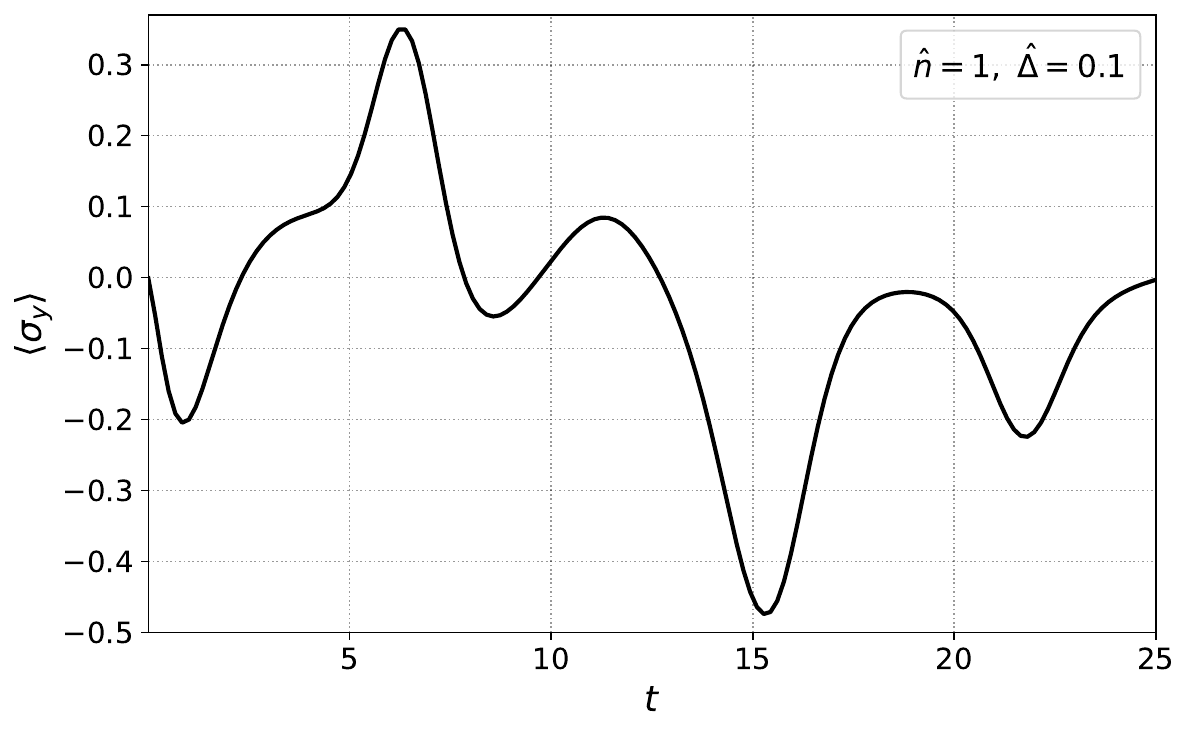}
    \includegraphics[width=0.235\textwidth]{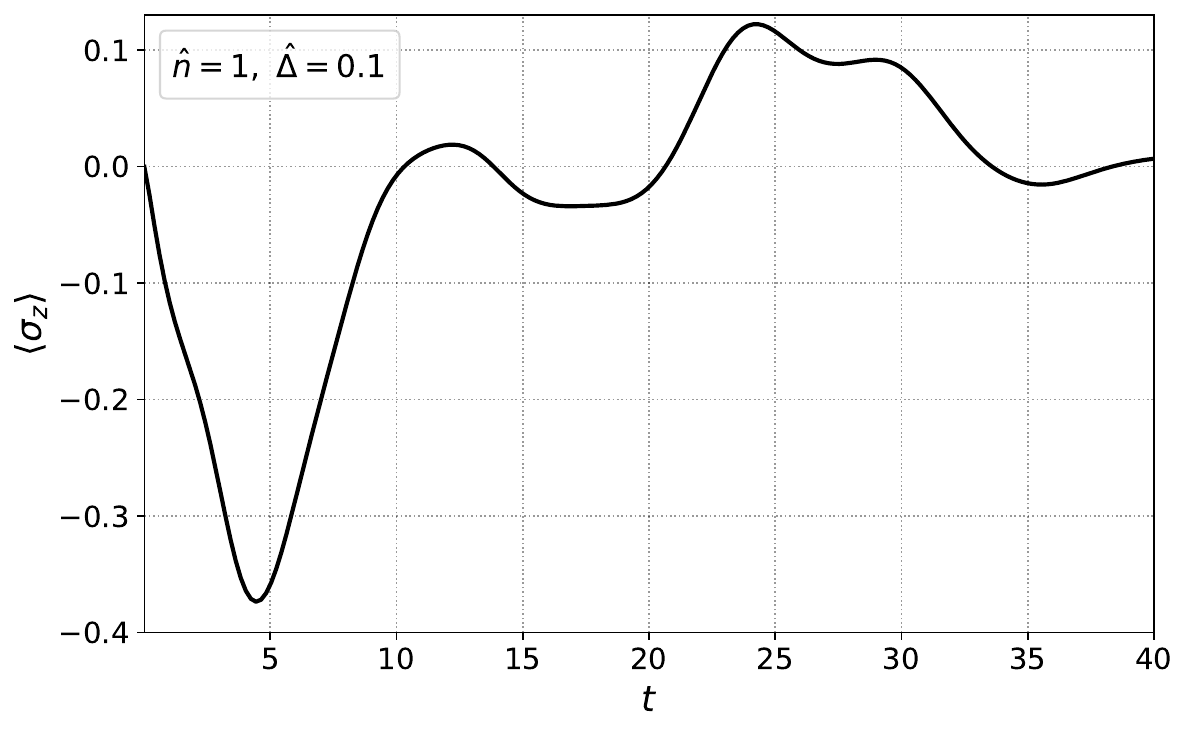}
	\caption{Spin observables in two-dimensional Bose gas at density $\hat{n}=1$ and Rabi coupling $\hat{\Delta}=0.1$.}
	\label{fig:2D_n=1_D=0.1}
\end{figure}

\begin{figure}[h!]
	\includegraphics[width=0.235\textwidth]{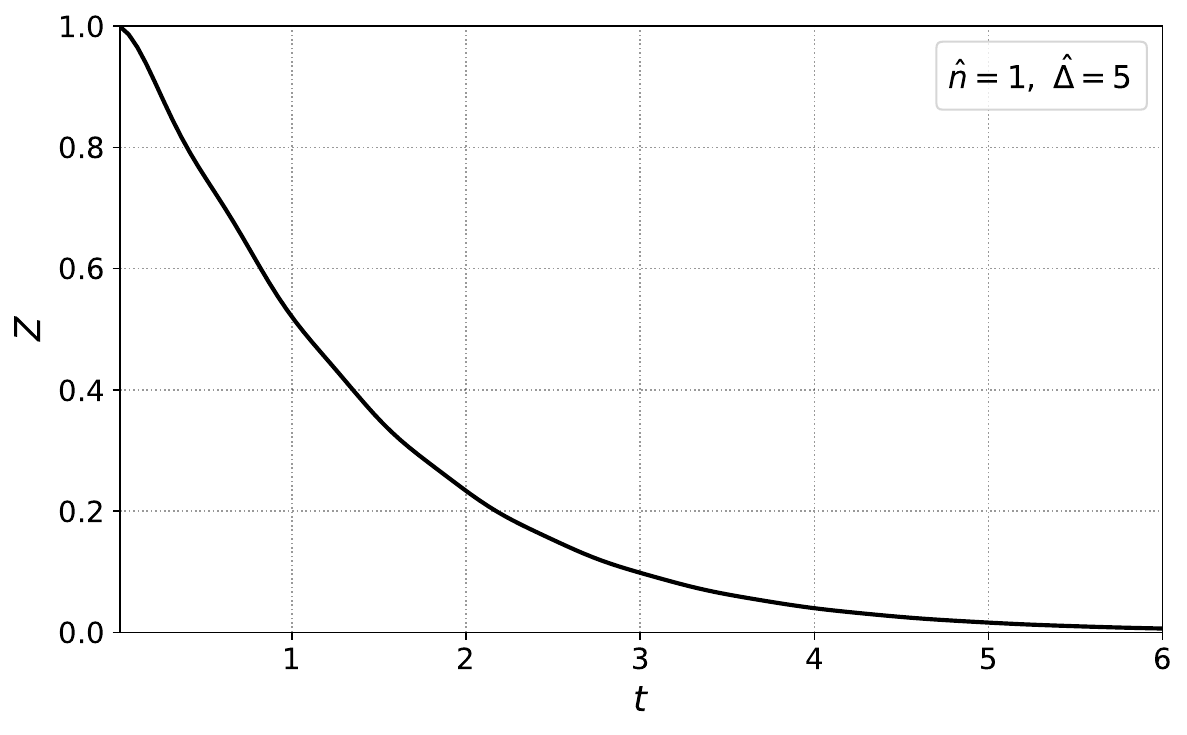}
    \includegraphics[width=0.235\textwidth]{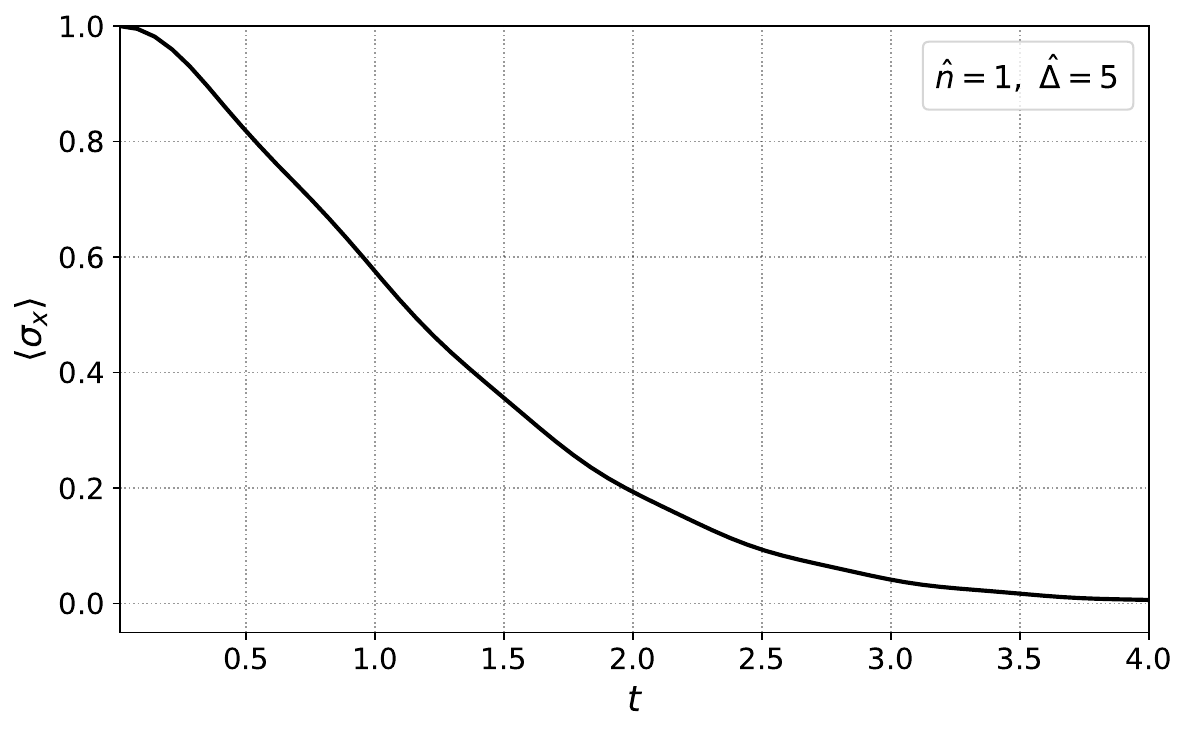}
    \includegraphics[width=0.235\textwidth]{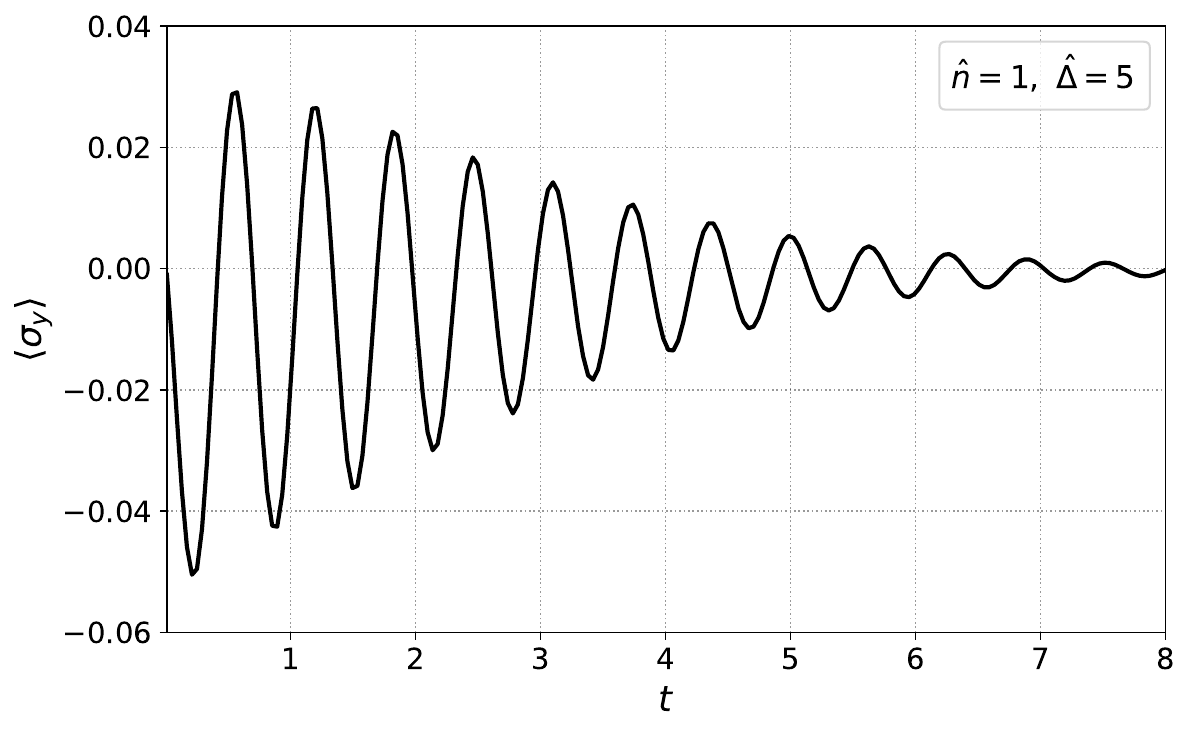}
    \includegraphics[width=0.235\textwidth]{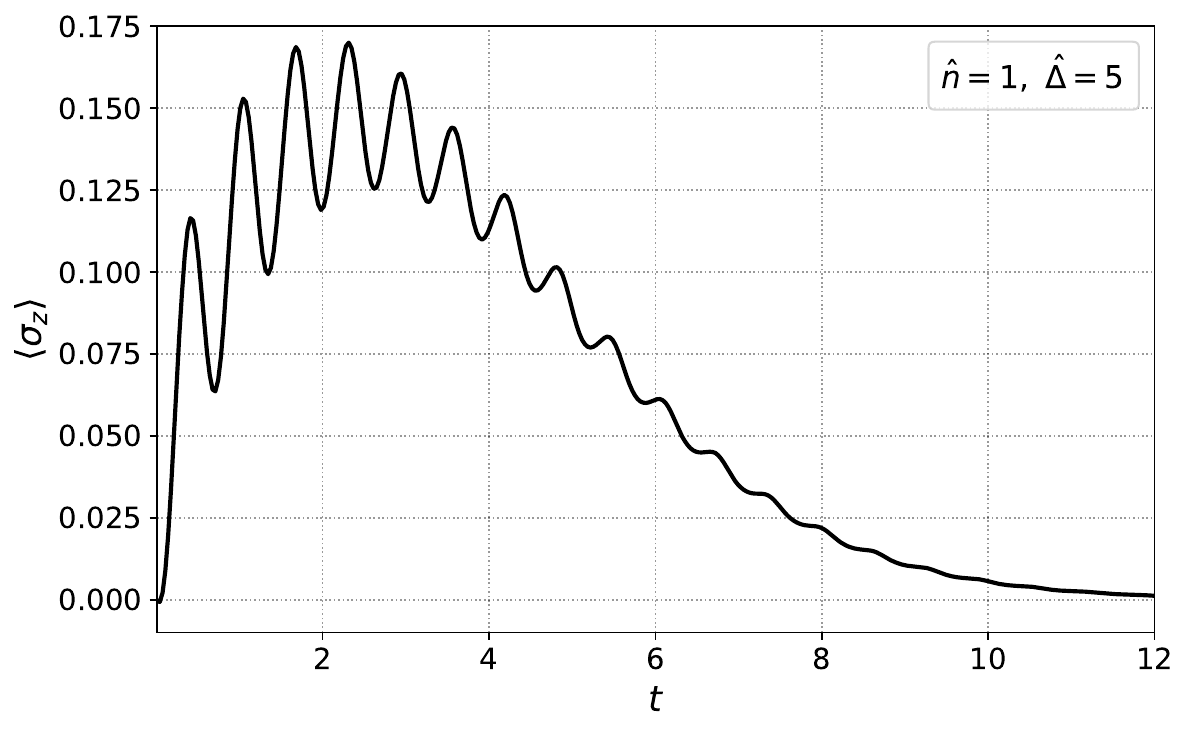}
	\caption{Similar density $\hat{n}=1$ but substantially larger transverse field $\hat{\Delta}=5$.}
	\label{fig:2D_n=1_D=5}
\end{figure}

In two spatial dimensions, the dimensionless coupling (or boson density) is chosen as follows: $\hat{n}=\frac{2\pi n}{m|\epsilon_{\uparrow}|}$, while notation for Rabi coupling $\hat{\Delta}=\Delta/|\epsilon_{\uparrow}|$ is preserved from the 3D case. The numerical procedure is similar to the previous one, with the formulas rewritten for the 2D case (the main changes are that the 3D boson-impurity $T$-matrix and the quantity $\Pi(\omega)$ are replaced by their 2D counterparts).
\begin{figure}[h!]
	\includegraphics[width=0.235\textwidth]{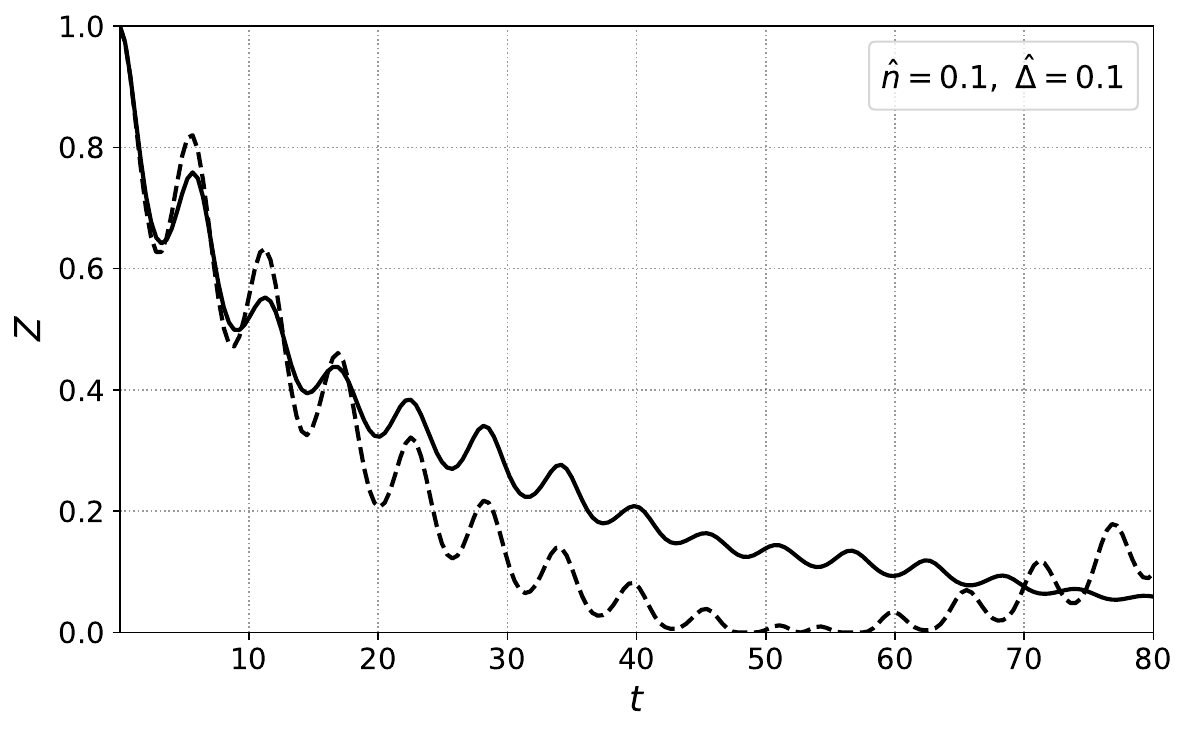}
    \includegraphics[width=0.235\textwidth]{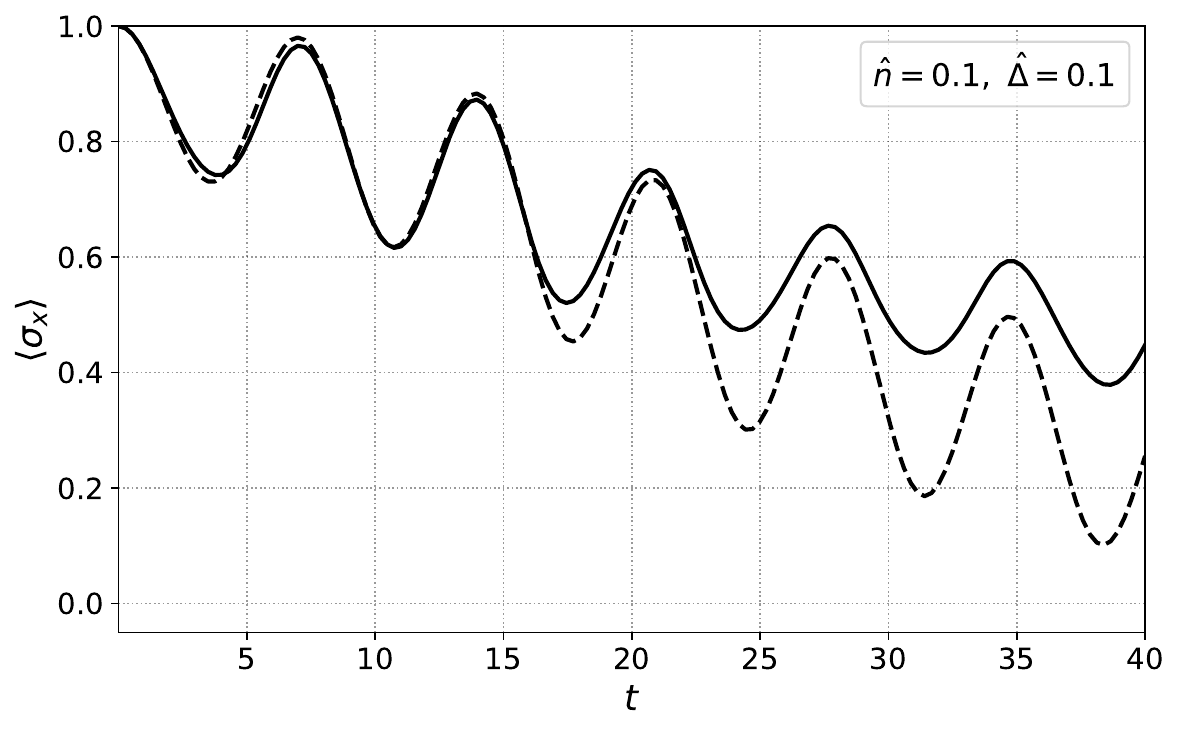}
    \includegraphics[width=0.235\textwidth]{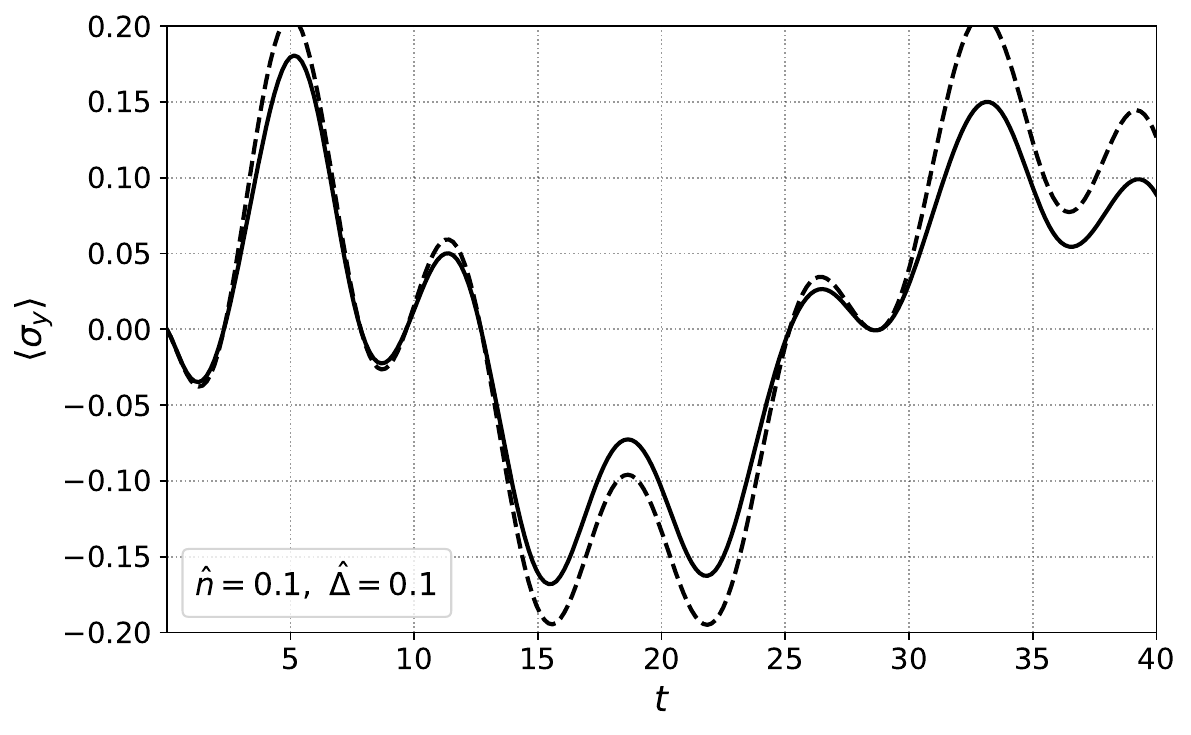}
    \includegraphics[width=0.235\textwidth]{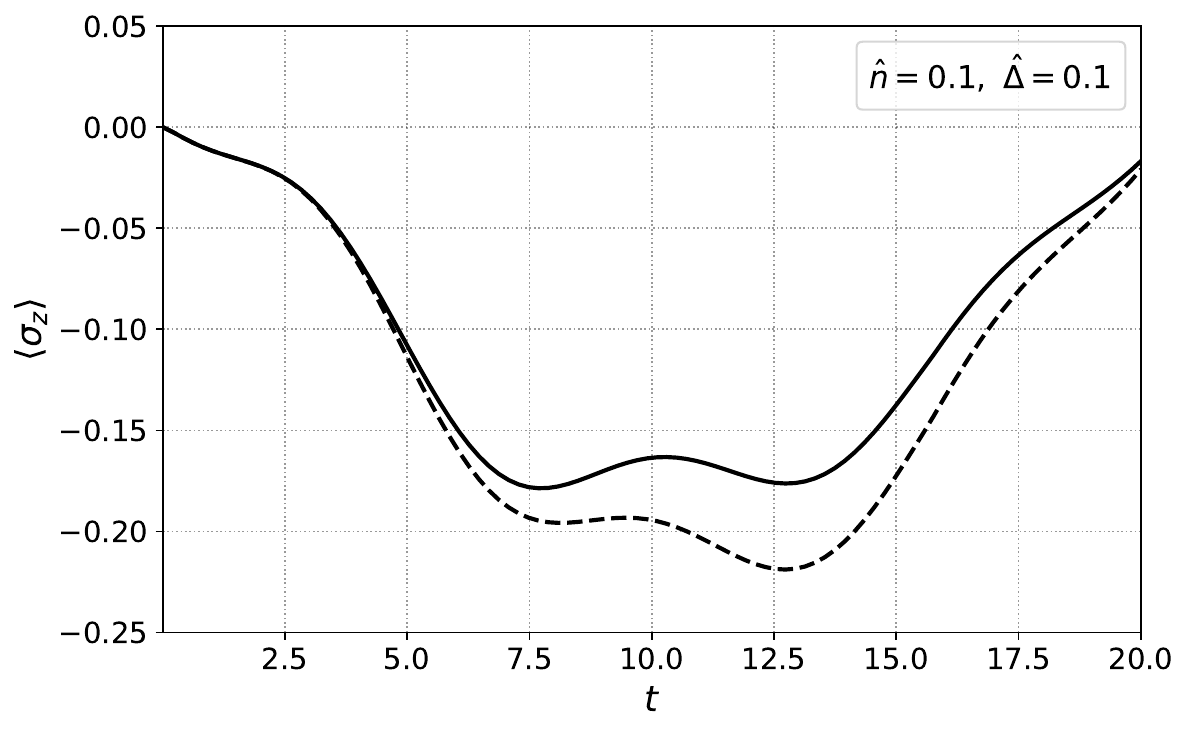}
	\caption{Two-dimensional time evolution of the modulus squared wave-function overlap $Z(t)$ and average spin components $\langle\sigma_{\alpha}\rangle$ for set of parameters $\hat{n}=0.1$ and $\hat{\Delta}=0.1$. Solid and dashed lines correspond to the results of our mean-field-like calculations and the linear-in-density-of-bosons approximation, respectively.}
	\label{fig:2D_n=0.1_D=0.1}
\end{figure}
\begin{figure}[h!]
	\includegraphics[width=0.235\textwidth]{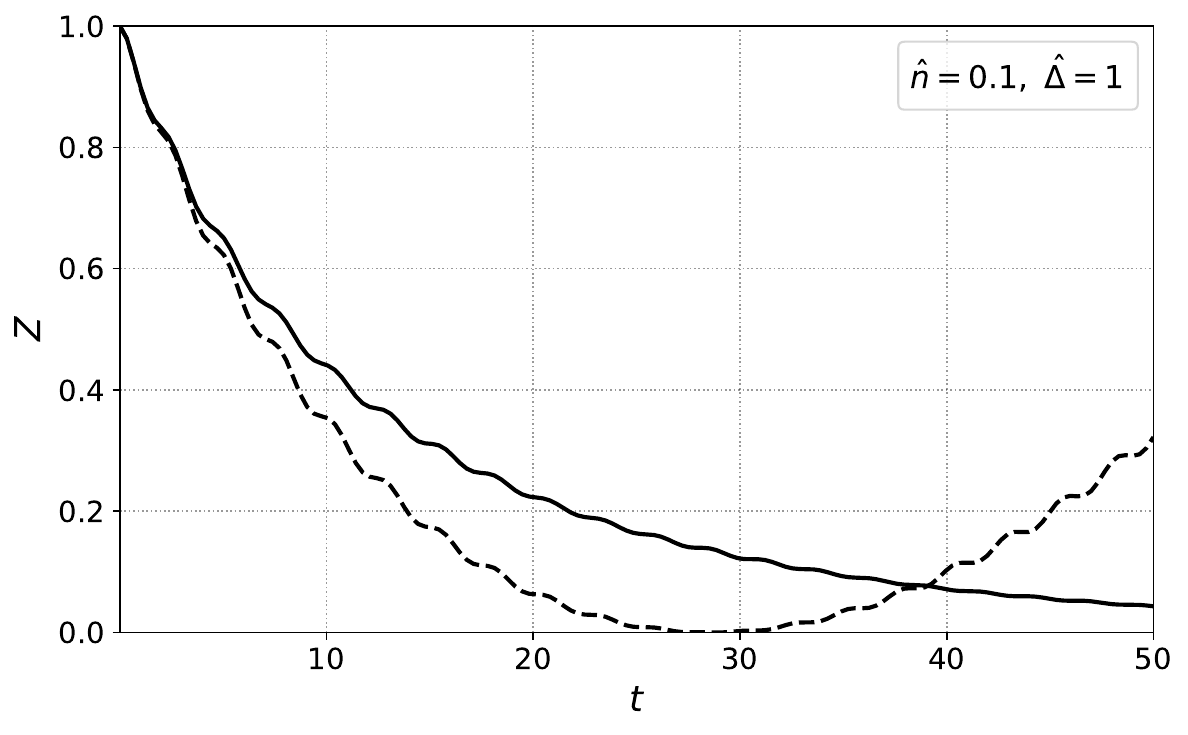}
    \includegraphics[width=0.235\textwidth]{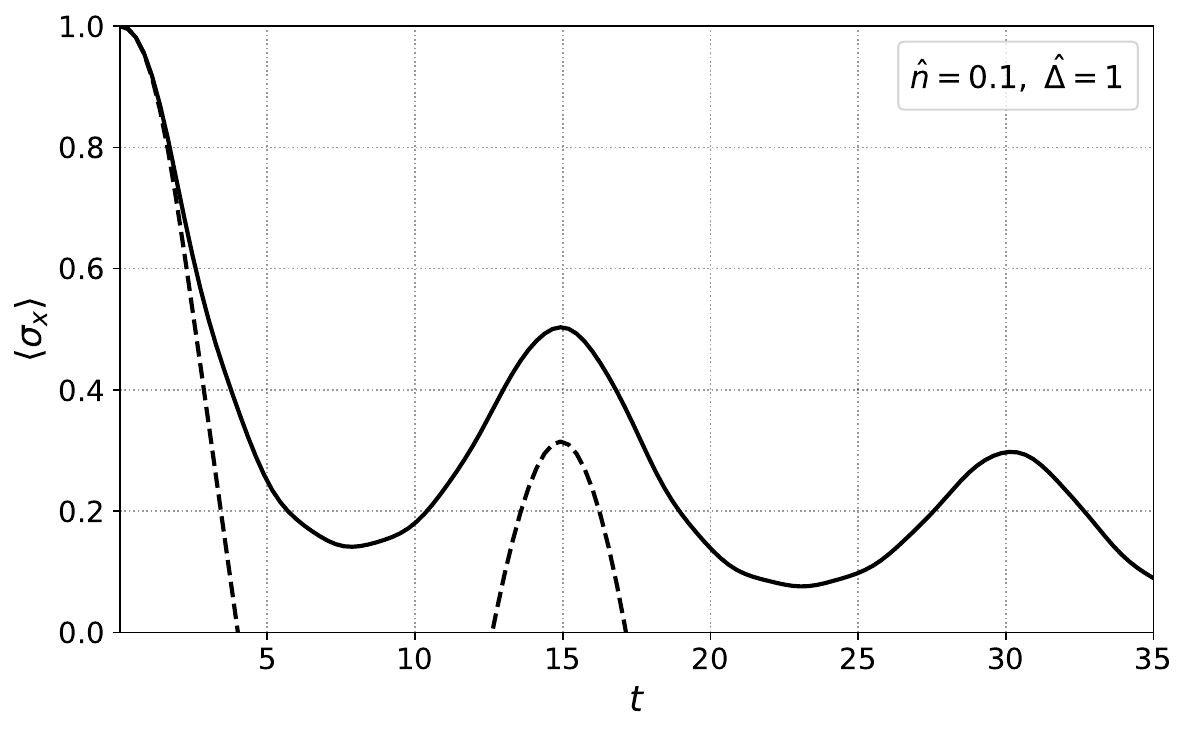}
    \includegraphics[width=0.235\textwidth]{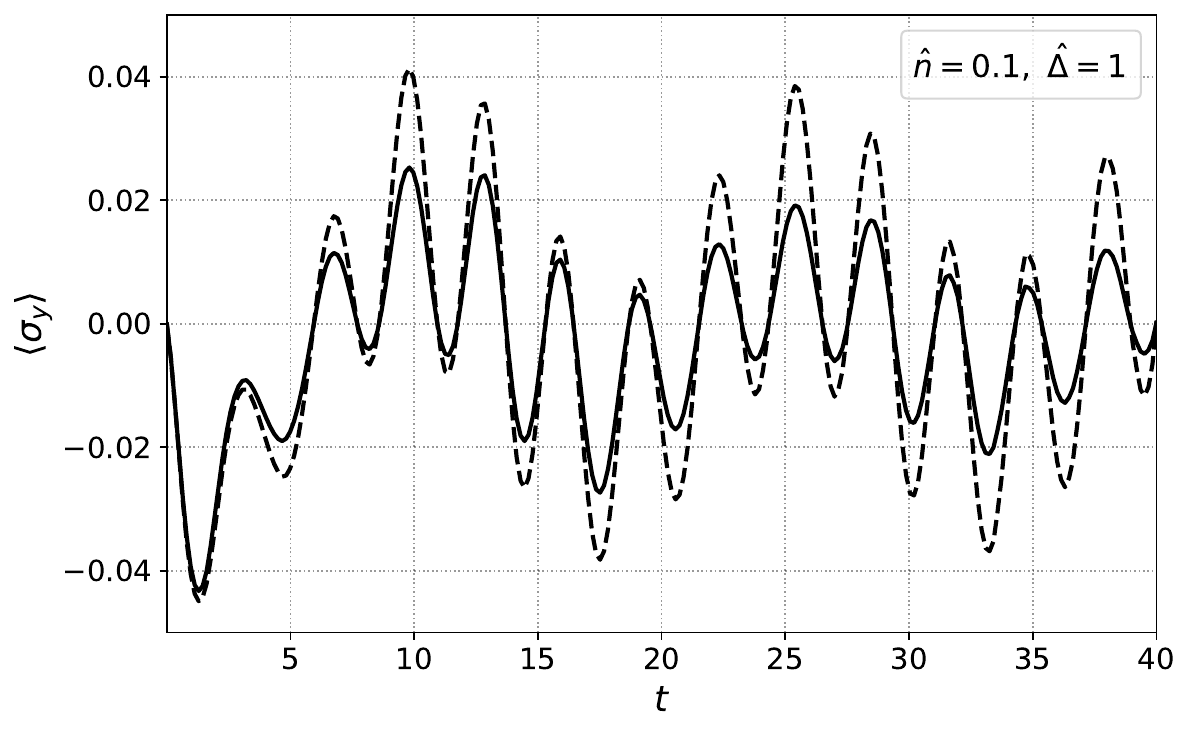}
    \includegraphics[width=0.235\textwidth]{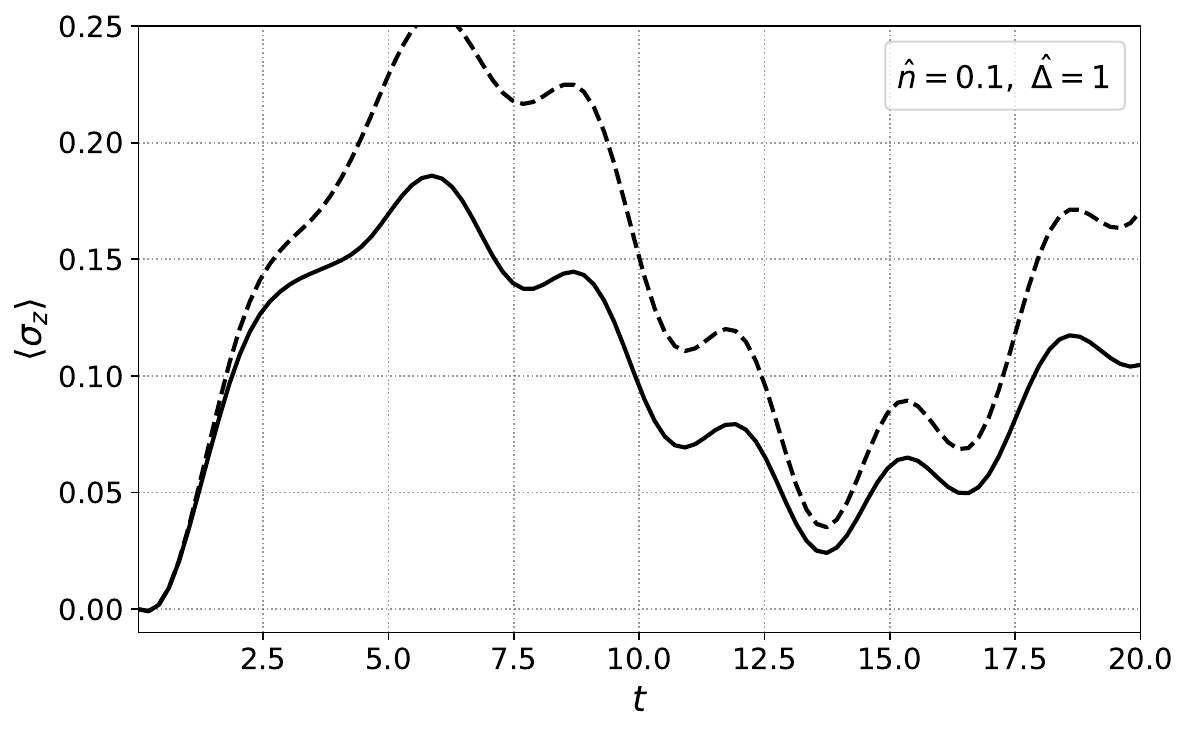}
	\caption{Same dimensionless density of bosons as in Fig.~\ref{fig:2D_n=0.1_D=0.1} $\hat{n}=0.1$ but stronger Rabi field $\hat{\Delta}=1$.}
	\label{fig:2D_n=0.1_D=1}
\end{figure}
At low densities of bosons (or weak boson-impurity coupling) $\hat{n}=0.1$, we compared (see Figs.~\ref{fig:2D_n=0.1_D=0.1}, \ref{fig:2D_n=0.1_D=1}) the results obtained within our mean-field calculations to the linear-in-density approximation. It is seen that an increase in the Rabi field leads to poorer consistency between the two curves. Larger transverse magnetic fields do not change the observed picture significantly; however, they require much more computational time.
As in the 3D case, and regardless of the magnitude of Rabi coupling, at low densities of bosons, we observe weak hints for steady state formation. They become more visible as the boson-impurity coupling increases. Figures~\ref{fig:2D_n=1_D=0.1} and \ref{fig:2D_n=1_D=5}, where the dimensionless density is taken as $\hat{n}=1$, support these conclusions.
The parameter sets were deliberately chosen to justify our approximate mean-field calculations. In general, the behavior of the two-dimensional system is qualitatively similar to its three-dimensional counterpart: an increase in the two-body interaction between bosons and spin leads to earlier thermalization of the system. While the latter conclusion is somewhat obvious, the stimulation of the steady-state formation by a transverse magnetic field is a less trivial fact. We believe this observation is due to the decreasing role of the boson-impurity bound state in shaping the system's dynamics. Although the value of the binding energy between the spin and bosons both in two and three dimensions increases for stronger Rabi couplings, its impact decreases due to the decrease of the bound-state residue. The latter turns zero at $\hat{\Delta}=2$ (resonant magnitude) in 3D and continuously decreases with a leading order power law asymptotics $1/\hat{\Delta}^2$ ($\hat{\Delta}\gg 1$) in 2D. Given the dominant role of the scattering states, the acceleration of spin thermalization appears natural.

\section{Summary}
In conclusion, we have studied the quench dynamics of a single spinful impurity in a gas of non-interacting bosons. In particular, we have considered a model with a nonzero boson-impurity interaction only for the spin-up state and a transverse Rabi field oriented along the $x$-axis. The system is exactly solvable in the few-boson limit, and not being in a position to solve the problem with a macroscopic number of Bose atoms due to the non-commutativity of spin terms in the Hamiltonian, we have proposed a mean-field-like approximation referring to a single-boson-impurity problem. While calculating the time dependence of various observables, the bosonic part of the full wave function is treated as a product of single-particle orbitals. This clearly contrasts with previous studies \cite{Liu_2026} in which the initial ground-state wave function is perturbed by a term with at most a single excitation. In our approach, all bosons evolve simultaneously, incorporating an arbitrary number of excited states. In two limiting cases, when the two-body interaction between bosons and the impurity, or the Rabi field, is separately absent, our calculations become exact. 

We have considered the two- and three-dimensional setups with the bosons prepared in a Bose-Einstein condensate and spin in the eigenstate of the $\sigma_x$ operator with eigenvalue $+1$ (this is the ground state of non-interacting subsystems); and the spin-dependent boson-impurity interaction suddenly switched on. Generally, the dynamical properties of the system in the two cases are qualitatively similar; nonetheless, at the single-boson-impurity level, there are some differences: the two-body bound state survives at any strength of the Rabi field in two dimensions. As expected, the stronger interaction with bosons produces faster spin decoherence (thermalization). This clearly tracks up by calculating the wave-function overlap between the initial state and the state at an arbitrary time moment, manifesting the orthogonality catastrophe, and by the decreasing dynamics of the averaged $\sigma_x$ from $+1$ to $0$. The steady state formation is also accompanied by the reduction of the amplitude of the intrinsic Rabi oscillations in the behavior of $\langle\sigma_y\rangle$ and $\langle\sigma_z\rangle$. Within the proposed approximation scheme, which more naturally accounts for collective effects in a bosonic medium, we observed a substantial speed-up in the amplitudes decay processes.

There are several directions for extending the proposed mean-field-like approach. The first is the inclusion of weak repulsion between bosons, which guarantees nonzero medium compressibility and can drastically affect the spin evolution. Finite-temperature calculations of the system's dynamics should provide a closer connection to experiments. Incorporating three-body effects into our mean-field treatment should shed light on the impact of Efimov physics on spin dynamics. Finally, it would be interesting to apply the approximation ideology incorporated in the mass-gap formulation of a fermionic bath.

\begin{center}
		{\bf Acknowledgments}
\end{center}
The work of O.H. was supported by Project 2025.07/0326 (No.~0126U002943) from the National Research Foundation of Ukraine.



\begin{thebibliography}{99}

\bibitem{Mistakidis_2023} S. I. Mistakidis, A. G. Volosniev, R. E. Barfknecht, T. Fogarty, Th. Busch, A. Foerster, P. Schmelcher, N. T. Zinner, {\it Few-body Bose gases in low dimensions--A laboratory for quantum dynamics.} \href{https://doi.org/10.1016/j.physrep.2023.10.004}{Phys. Rep. {\bf 1042}, 1 (2023).}
\bibitem{Grusdt_2025} F. Grusdt, N. Mostaan, E. Demler, L. A. Pe\~na Ardila, {\it Impurities and polarons in bosonic quantum gases: a review on recent progress.} \href{https://doi.org/10.1088/1361-6633/add94b}{Rep. Prog. Phys. {\bf 88}, 066401 (2025).}
\bibitem{Massignan_2026} P. Massignan, R. Schmidt, G. E. Astrakharchik, A. Imamoglu, M. Zwierlein, J. J. Arlt, G. M. Bruun, {\it Polarons in atomic gases and two-dimensional semiconductors.} \href{https://doi.org/10.1103/4nng-bb9z}{Rev. Mod. Phys. {\bf 98}, 035002 (2026).}



\bibitem{Grusdt_2016} F. Grusdt and E. A. Demler, {\it New Theoretical Approaches to Bose Polarons.} edited by M. Inguscio, W. Ketterle, S.
Stringari, and G. Roati (IOS PRESS, Amsterdam, 2016),
p. 325.
\bibitem{Grusdt_2015} F. Grusdt, Y. E. Shchadilova, A. N. Rubtsov, and E. Demler {\it Renormalization group approach to the Fr\"ohlich polaron model: application to impurity-BEC problem.} \href{https://doi.org/10.1038/srep12124}{Sci. Rep. {\bf 5}, 12124 (2015).}
\bibitem{Shchadilova_2016} Y. E. Shchadilova, F. Grusdt, A. N. Rubtsov, and E. Demler, {\it Polaronic mass renormalization of impurities in Bose-Einstein condensates: Correlated Gaussian-wave-function approach.} \href{http://doi.org/10.1103/PhysRevA.93.043606}{Phys. Rev. A {\bf 93}, 043606 (2016).}


\bibitem{Frohlich_1954} H. Fr\"ohlich, {\it Electrons in lattice fields.} \href{https://doi.org/10.1080/00018735400101213}{Adv. Phys. {\bf 3}, 325 (1954).}

\bibitem{Grusdt_2017} F. Grusdt, G. E. Astrakharchik, and E. A. Demler, {\it Bose polarons in ultracold atoms in one dimension: beyond the Fr\" ohlich paradigm.} \href{https://doi.org/10.1088/1367-2630/aa8a2e}{New J. Phys. {\bf 19}, 103035 (2017).}
\bibitem{Kain_2018} B. Kain and H. Y. Ling, {\it Analytical study of static beyond-Fr\"ohlich Bose polarons in one dimension.} \href{ https://doi.org/10.1103/PhysRevA.98.033610}{Phys. Rev. A {\bf 98}, 033610 (2018).}
\bibitem{Ichmoukhamedov_2019} T. Ichmoukhamedov and J. Tempere,
{\it Feynman path-integral treatment of the Bose polaron beyond the Fr\"ohlich model.} \href{https://doi.org/10.1103/PhysRevA.100.043605}{Phys. Rev. A {\bf 100}, 043605 (2019).}
\bibitem{Christ_2025} J.-P. Christ, P. Bermes, and F. Grusdt, {\it Operator-valued-flow-equation approach to the bosonic lattice polaron: Dispersion renormalization beyond the Fr\"ohlich paradigm.} \href{https://doi.org/10.1103/pcsy-czjp}{Phys. Rev. A {\bf 112}, 033317 (2025).}

\bibitem{Rath_2013} S. P. Rath and R. Schmidt, {\it Field-theoretical study of the Bose polaron.} \href{https://doi.org/10.1103/PhysRevA.88.053632}{Phys. Rev. A {\bf 88}, 053632 (2013).}
\bibitem{Li_2014} W. Li and S. D. Sarma, {\it Variational study of polarons in Bose-Einstein condensates.} \href{https://doi.org/10.1103/PhysRevA.90.013618}{Phys. Rev. A {\bf 90}, 013618 (2014).}
\bibitem{Pena_Ardila_2015} L. A. Pe\~na Ardila and S. Giorgini, {\it Impurity in a Bose-Einstein condensate: Study of the attractive and repulsive branch using quantum Monte Carlo methods.} \href{https://doi.org/10.1103/PhysRevA.92.033612}{Phys. Rev. A {\bf 92}, 033612 (2015).}
\bibitem{Alhyder_2026} R. Alhyder, G. M. Bruun, T. Pohl, M. Lemeshko, and A. G. Volosniev, {\it Phenomenological model of decaying Bose polarons.} \href{https://doi.org/10.1103/16dk-5dgx}{Phys. Rev. Research {\bf 8}, L012034 (2026).}


\bibitem{Hu_2016} M.-G. Hu, Michael J. Van de Graaff, D. Kedar, J. P. Corson, E. A. Cornell, and D. S. Jin {\it Bose Polarons in the Strongly Interacting Regime.} \href{https://doi.org/10.1103/PhysRevLett.117.055301}{Phys. Rev. Lett. {\bf 117}, 055301 (2016).}
\bibitem{Jorgensen_2016} N. B. J\o{}rgensen, L. Wacker, K. T. Skalmstang, M. M. Parish, J. Levinsen, R. S. Christensen, G. M. Bruun, and J. J. Arlt, {\it Observation of Attractive and Repulsive Polarons in a Bose-Einstein Condensate.} \href{https://doi.org/10.1103/PhysRevLett.117.055302}{Phys. Rev. Lett. {\bf 117}, 055302 (2016).}
\bibitem{Yan_2020} Z. Z. Yan, Y. Ni, C. Robens, and M. W. Zwierlein, {\it Bose polarons near quantum criticality.} \href{https://doi.org/10.1126/science.aax5850}{Science {\bf 368}, 190 (2020).}
\bibitem{Skou_2022} M. G. Skou, K. K. Nielsen, T. G. Skov, A. M. Morgen, N. B.
J\o{}rgensen, A. Camacho-Guardian, T. Pohl, G. M. Bruun,
and J. J. Arlt, {\it Life and death of the Bose polaron.} \href{https://doi.org/10.1103/PhysRevResearch.4.043093}{Phys. Rev.
Res. {\bf 4}, 043093 (2022).}
\bibitem{Etrych_2025} J. Etrych, G. Martirosyan, A. Cao, C. J. Ho, Z. Hadzibabic, and C. Eigen, {\it Universal Quantum Dynamics of Bose Polarons.} \href{https://doi.org/10.1103/PhysRevX.15.021070}{Phys. Rev. X {\bf 15}, 021070 (2025).}

\bibitem{Yoshida_2018} S. M. Yoshida, Z.-Y. Shi, J. Levinsen, M. M. Parish, {\it Few-body states of bosons interacting with a heavy quantum impurity.} \href{https://doi.org/10.1103/PhysRevA.98.062705}{Phys. Rev. A {\bf 98}, 062705 (2018).}
\bibitem{Shi_2018} Z.-Y. Shi, S. M. Yoshida, M. M. Parish, J. Levinsen, {\it Impurity-Induced Multibody Resonances in a Bose Gas.} \href{https://doi.org/10.1103/PhysRevA.98.062705}{Phys. Rev. Lett. {\bf 121}, 243401 (2018).}
\bibitem{Levinsen_2021} J. Levinsen, L. A. Pe\~na Ardila, S. M. Yoshida, and M. M. Parish, {\it Quantum Behavior of a Heavy Impurity Strongly Coupled to a Bose Gas.} \href{https://doi.org/10.1103/PhysRevLett.127.033401}{Phys. Rev. Lett. {\bf 127}, 033401 (2021).}


\bibitem{Efimov_1970} V. Efimov, {\it Energy levels arising from resonant two-body forces in a three-body system.} \href{https://doi.org/10.1016/0375-9474(73)90510-1}{Phys. Lett. B {\bf 33}, 563 (1970).}

\bibitem{Levinsen_2015} J. Levinsen, M. M. Parish, and G. M. Bruun, {\it Impurity in a Bose-Einstein Condensate and the Efimov Effect.} \href{https://doi.org/10.1103/PhysRevLett.115.125302}{Phys. Rev. Lett. {\bf 115}, 125302 (2015).}
\bibitem{Sun_2017} M. Sun, H. Zhai, and X. Cui, {\it Visualizing the Efimov Correlation in Bose Polarons.} \href{https://doi.org/10.1103/PhysRevLett.119.013401}{Phys. Rev. Lett. {\bf 119}, 013401 (2017).}
\bibitem{Yoshida_2018_2} S. M. Yoshida, S. Endo, J. Levinsen, and M. M. Parish, {\it Universality of an Impurity in a Bose-Einstein Condensate.} \href{https://doi.org/10.1103/PhysRevX.8.011024}{Phys. Rev. X 8, 011024 (2018).}
\bibitem{Blume_2019} D. Blume, {\it Few-boson system with a single impurity: Universal bound states tied to Efimov trimers.} \href{https://doi.org/10.1103/PhysRevA.99.013613}{Phys. Rev. A {\bf 99}, 013613 (2019).}

\bibitem{Christianen_2022} A. Christianen, J. I. Cirac, and R. Schmidt, {\it Bose polaron and the Efimov effect: A Gaussian-state approach.} \href{https://doi.org/10.1103/PhysRevA.105.053302}{Phys. Rev. A {\bf 105}, 053302 (2022).}
\bibitem{Christianen_2022_2} A. Christianen, J. I. Cirac, and R. Schmidt, {\it Chemistry of a Light Impurity in a Bose-Einstein Condensate.} \href{https://doi.org/10.1103/PhysRevLett.128.183401}{Phys. Rev. Lett. {\bf 128}, 183401 (2022).}

\bibitem{Mostaan_2025} N. Mostaan, N. Goldman, and F. Grusdt, {\it Unified theory of strong coupling Bose polarons: From repulsive polarons to non-Gaussian many-body bound states.} \href{https://doi.org/10.1103/7vdd-8vb4}{Phys. Rev. Research {\bf 7}, 043349 (2025).}


\bibitem{Robens_2026} C. Robens, A. Christianen, A. Y. Chuang, H. Q. Bui, Y. Zhang, R. Schmidt, M. Zwierlein, {\it Polaronic hybridization of atoms, dimers and trimers in a Bose-Einstein condensate.} \href{https://doi.org/10.48550/arXiv.2606.20440}{arXiv:2606.20440}



\bibitem{Skou_2021} M. G. Skou, T. G. Skov, N. B. J\o{}rgensen, K. K. Nielsen, A. Camacho-Guardian, T. Pohl, G. M. Bruun, and J. J. Arlt, {\it Non-equilibrium quantum dynamics and formation of the Bose
polaron.} \href{https://doi.org/10.1038/s41567-021-01184-5}{Nat. Phys. {\bf 17}, 731 (2021).}
\bibitem{Shchadilova_2016_2} Y. E. Shchadilova R. Schmidt, F. Grusdt, E. Demler, {\it Quantum Dynamics of Ultracold Bose Polarons.} \href{https://doi.org/10.1103/PhysRevLett.117.113002}{Phys. Rev. Lett. {\bf 117}, 113002 (2016).}
\bibitem{Nielsen_2019} K. K. Nielsen, L. A. Pe\~na Ardila, G. M. Bruun and T. Pohl, {\it Critical slowdown of non-equilibrium polaron dynamics.} \href{https://doi.org/10.1088/1367-2630/ab0a81}{New J. Phys. {\bf 21}, 043014 (2019).}
\bibitem{Dzsotjan_2020} D. Dzsotjan, R. Schmidt, and M. Fleischhauer,
{\it Dynamical Variational Approach to Bose Polarons at Finite Temperatures.}
\href{https://doi.org/10.1103/PhysRevLett.124.223401}{Phys. Rev. Lett. {\bf 124}, 223401 (2020).}


\bibitem{Volosniev_2015} A. G. Volosniev, H.-W. Hammer, and N. T. Zinner, {\it Real-time dynamics of an impurity in an ideal Bose gas in a trap.} \href{https://doi.org/10.1103/PhysRevA.92.023623}{Phys. Rev. A {\bf 92}, 023623 (2015).}
\bibitem{Field_2020} B. Field, J. Levinsen, and M. M. Parish, {\it Fate of the Bose polaron at finite temperature.} \href{https://doi.org/10.1103/PhysRevA.101.013623}{Phys. Rev. A {\bf 101}, 013623 (2020).}
\bibitem{Pena_Ardila_2021} L. A. Pe\~na Ardila, {\it Dynamical formation of polarons in a Bose-Einstein condensate: A variational approach.} \href{https://doi.org/10.1103/PhysRevA.103.033323}{Phys. Rev. A {\bf 103}, 033323 (2021).}
\bibitem{Drescher_2021} M. Drescher, M. Salmhofer, and T. Enss, {\it Quench Dynamics of the Ideal Bose Polaron at Zero and Nonzero Temperatures.} \href{https://doi.org/10.1103/PhysRevA.103.033317}{Phys. Rev. A {\bf 103}, 033317 (2021).}

\bibitem{Shashi_2014} A. Shashi, F. Grusdt, D. Abanin, and E. Demler, {\it Radio-frequency spectroscopy of polarons in ultracold Bose gases.} \href{https://doi.org/10.1103/PhysRevA.89.053617}{Phys. Rev. A {\bf 89}, 053617 (2014).}
\bibitem{Adam_2022} D. Adam, Q. Bouton, J. Nettersheim, S. Burgardt, and A. Widera, {\it Coherent and Dephasing Spectroscopy for Single-Impurity Probing of an Ultracold Bath.} \href{https://doi.org/10.1103/PhysRevLett.129.120404}{Phys. Rev. Lett. {\bf 129}, 120404 (2022).}


		
\bibitem{Mistakidis_2019} S. I. Mistakidis, G. C. Katsimiga, G. M. Koutentakis, T. Busch, and P. Schmelcher, {\it Quench Dynamics and Orthogonality Catastrophe of Bose Polarons.} \href{https://doi.org/10.1103/PhysRevLett.122.18300}{Phys. Rev. Lett. {\bf 122}, 183001 (2019).}
\bibitem{Hryhorchak_2026} O. Hryhorchak, G. Panochko, and V. Pastukhov, {\it Evolution of a single spin in ideal Bose gas at finite temperatures.} \href{https://doi.org/10.1088/1751-8121/ae7229}{J. Phys. A: Math. Theor. {\bf 59}, 225304 (2026).}

\bibitem{Schmidt_2016} R. Schmidt, H. R. Sadeghpour, and E. Demler, {\it Mesoscopic Rydberg Impurity in an Atomic Quantum Gas.} \href{https://doi.org/10.1103/PhysRevLett.116.105302}{Phys. Rev. Lett. {\bf 116}, 105302 (2016).}
\bibitem{Camargo_2018} F. Camargo {\it et al.}, {\it Creation of Rydberg Polarons in a Bose Gas.} \href{https://doi.org/10.1103/PhysRevLett.120.083401}{Phys. Rev. Lett. {\bf 120}, 083401 (2018).}
\bibitem{Durst_2024} A. A. T. Durst and M. T. Eiles, {\it Phenomenology of a Rydberg impurity in an ideal Bose-Einstein condensate.} \href{https://doi.org/10.1103/PhysRevResearch.6.L042009}{Phys. Rev. Research {\bf 6}, L042009 (2024).}



\bibitem{Mulkerin_2024} B. C. Mulkerin, J. Levinsen, and Meera M. Parish, {\it Rabi oscillations and magnetization of a mobile spin-1/2 impurity in a Fermi sea.} \href{https://doi.org/10.1103/PhysRevA.109.023302}{Phys. Rev. A {\bf 109}, 023302 (2024).}
\bibitem{Wasak_2024} T. Wasak, M. Sighinolfi, J. Lang, F. Piazza, and A. Recati, {\it Decoherence and Momentum Relaxation in Fermi-Polaron Rabi Dynamics: A Kinetic Equation Approach.} \href{https://doi.org/10.1103/PhysRevLett.132.183001}{Phys. Rev. Lett. {\bf 132}, 183001 (2024).}
\bibitem{Bleu_2025} O. Bleu, B. C. Mulkerin, C. R. Cabrera, J. Levinsen, and M. M. Parish, {\it Scattering resonances and pairing in a Rabi-coupled Fermi gas.} \href{https://doi.org/10.1103/stvf-cn1b}{Phys. Rev. A {\bf 112}, L011304 (2025).}
\bibitem{Vivanco_2025} Vivanco, F. J., Schuckert, A., Huang, S., Schumacher, G. L., Assumpcao, G. G., Ji, Y., \dots Navon, N. {\it The strongly driven Fermi polaron.} \href{https://doi.org/10.1038/s41567-025-02799-8}{Nat. Phys. {\bf 21}, 564 (2025).}

\bibitem{Liu_2026} Zeyu Liu and Pengfei Zhang, {\it Rabi oscillations of strongly driven Bose polarons.} \href{https://doi.org/10.1103/s13m-54d7}{Phys. Rev. Research {\bf 8}, L012010 (2026).}

\bibitem{Hryhorchak_2023} O. Hryhorchak and V. Pastukhov, {\it Second root of dilute Bose-Fermi mixtures.} \href{https://doi.org/10.1088/1751-8121/accda4}{J. Phys. A: Math. Theor. {\bf 56}, 205003 (2023).}

\bibitem{Panochko_2021} G. Panochko and V. Pastukhov, {\it Two- and three-body effective potentials between impurities in ideal BEC.} \href{https://doi.org/10.1088/1751-8121/abdbc5}{J. Phys. A: Math. Theor. {\bf 54}, 085001 (2021).}
\bibitem{Panochko_2022} G. Panochko and V. Pastukhov, {\it Static Impurities in a Weakly Interacting Bose Gas.} \href{https://doi.org/10.3390/atoms10010019}{Atoms {\bf 10}, 19 (2022).}
\bibitem{Hryhorchak_2023_2} O. Hryhorchak and V. Pastukhov, {\it Trapped Ideal Bose Gas with a Few Heavy Impurities.} \href{https://doi.org/10.3390/atoms11050077}{Atoms {\bf 11}, 77 (2023).}

\bibitem{LLP_1953} T. D. Lee, F. E. Low, and D. Pines, {\it The Motion of Slow Electrons in a Polar Crystal.} \href{https://doi.org/10.1103/PhysRev.90.297}{Phys. Rev. {\bf 90}, 297 (1953).}

\bibitem{Kain_2017} B. Kain and H. Y. Ling, {\it Hartree-Fock treatment of Fermi polarons using the Lee-Low-Pine transformation.} \href{}{Phys.
Rev. A {\bf 96}, 033627 (2017).}
\bibitem{Chen_2025} X. Chen, E. Dizer, E. R. Rodr\'{\i}guez, and R. Schmidt, {\it Mass-Gap Description of Heavy Impurities in Fermi Gases.}
\href{https://doi.org/10.1103/h2f7-dhjh}{Phys. Rev. Lett. {\bf 118}, 193401 (2025).}

\bibitem{Gross_1962} E. Gross, {\it Motion of foreign bodies in boson systems.} \href{https://doi.org/10.1016/0003-4916(62)90217-8}{Ann. Phys. {\bf 19}, 234 (1962).}
\bibitem{Hryhorchak_2020} O. Hryhorchak, G. Panochko, V. Pastukhov, {\it Mean-field study of repulsive 2D and 3D Bose polarons.} \href{https://doi.org/10.1088/1361-6455/abb3ab}{J. Phys. B At. Mol. Opt. Phys. {\bf 53}, 205302 (2020).}
\bibitem{Hryhorchak_2020_2} O. Hryhorchak, G. Panochko, V. Pastukhov, {\it Impurity in a three-dimensional unitary Bose gas.} \href{https://doi.org/10.1016/j.physleta.2020.126934}{Phys. Lett. A {\bf 384}, 126934 (2020).}
\bibitem{Massignan_2021} P. Massignan, N. Yegovtsev, V. Gurarie, {\it Universal Aspects of a Strongly Interacting Impurity in a Dilute Bose Condensate.} \href{https://doi.org/10.1103/PhysRevLett.126.123403}{Phys. Rev. Lett. {\bf 126}, 123403 (2021).} \href{https://doi.org/10.1103/PhysRevLett.126.123403}{Phys. Rev. Lett. {\bf 126}, 123403 (2021).}
\bibitem{Guenther_2021} N. E. Guenther, R. Schmidt, G. M. Bruun, V. Gurarie, P. Massignan, {\it Mobile impurity in a Bose-Einstein condensate and the orthogonality catastrophe.} \href{https://doi.org/10.1103/PhysRevA.103.013317}{Phys. Rev. A {\bf 103}, 013317 (2021).}
\bibitem{Schmidt_2022} R. Schmidt, T. Enss, {\it Self-stabilized Bose polarons.} \href{https://doi.org/10.21468/SciPostPhys.13.3.054}{SciPost Phys. {\bf 13}, 054 (2022).}

\bibitem{Volosniev_2017} A. G. Volosniev and H. W. Hammer, {\it Analytical approach to the Bose-polaron problem in one dimension.} \href{https://doi.org/10.1103/PhysRevA.96.031601}{Phys. Rev. A {\bf 96}, 031601 (2017).}
\bibitem{Pastukhov_2019} V. Pastukhov, {\it Mean-field properties of impurity in Bose gas with three-body forces.} \href{https://doi.org/10.1016/j.physleta.2019.05.018}{Phys. Lett. A {\bf 383}, 2610 (2019).}
\bibitem{Panochko_2019} G. Panochko and V. Pastukhov, {\it Mean-field construction for spectrum of one-dimensional Bose polaron.} \href{https://doi.org/10.1088/1751-8121/abdbc5}{Ann. Phys. {\bf 409}, 167933 (2019).}
\bibitem{Petkovic_2022} A. Petkovi\ifmmode \acute{c}\else \'{c}\fi{}, {\it Local spectral density of an interacting one-dimensional Bose gas with an impurity.} \href{https://doi.org/10.1103/PhysRevA.105.043305}{Phys. Rev. A {\bf 105}, 043305 (2022).}
\bibitem{Petkovic_2023} A. Petkovi\ifmmode \acute{c}\else \'{c}\fi{} and  Z. Ristivojevic, {\it Density of a one-dimensional weakly interacting Bose gas far from an impurity.} \href{https://doi.org/10.1103/PhysRevB.108.174510}{Phys. Rev. B {\bf 108}, 174510 (2023).}

\bibitem{Jager_2020} J. Jager, R. Barnett, M. Will, M. Fleischhauer, {\it Strong-coupling Bose polarons in one dimension: Condensate deformation and modified Bogoliubov phonons.} \href{https://doi.org/10.1103/PhysRevResearch.2.033142}{Phys. Rev. Res. {\bf 2}, 033142 (2020).}

\bibitem{Moccia_1973} R. Moccia, {\it Time-dependent variational principle}. \href{https://doi.org/10.1002/qua.560070414}{Int. J. Quantum Chem. {\bf 7}, 779 (1973).}

\end{thebibliography}
\end{document}